\documentclass[12pt]{iopart}

\pdfoutput=1

\usepackage{graphicx, cite}
\newcommand{\M}{\mathcal{M}}

\usepackage{color}

\def \be{\begin{equation}}
\def \ee{\end{equation}}
\def \bew{\begin{widetext}\begin{equation}}
\def \eew{\end{equation}\end{widetext}}
\def \bmlett{\begin{mathletters}}
\def \emlett{\end{mathletters}}

\def \nbar{n_{\rm eq} }

\def \II{{\mathcal I}}

\def \tn{\tilde{n}}
\def \tm{\tilde{m}}
\def \tP{\tilde{P}}

\def \ra{\rightarrow}

\def \tavg{t_{\rm avg}}
\def \Tbath{T_{\rm bath}}
\def \Tinit{T_{\rm init}}
\def \ninit{n_{\rm init}}
\def \Pmeas{P_{\rm meas}}
\def \Pmeascl{P_{\rm meas,cl}}
\def \Pmeasq{P_{\rm meas,q}}
\begin{document}

%\preprint{}

\title{Dispersive optomechanics: a membrane inside a cavity}

\author{A. M. Jayich$^1$, J. C. Sankey$^1$, B. M. Zwickl$^1$, C. Yang$^1$, J. D. Thompson$^1$, S. M. Girvin$^{1,2}$, A. A. Clerk$^3$, F. Marquardt$^4$, J. G. E. Harris$^{1,2}$}
%A. M. Jayich, J. C. Sankey, B. M. Zwickl, C. Yang, J. D. Thompson, S. M. Girvin, A. A. Clerk, F. Marquardt, J. G. E. Harris
%\thanks{}

\address{1. Department of Physics, Yale University, New Haven, CT}
\address{2. Department of Applied Physics, Yale University, New Haven, CT}
\address{3. Department of Physics, McGill University, Montreal, Canada}
\address{4. Department of Physics, Arnold-Sommerfeld-Center for Theoretical Physics and Center for Nanoscience, M\"unchen, Germany} 


\date{\today}

\begin{abstract}
We present the results of theoretical and experimental studies of dispersively coupled (or      
``membrane in the middle'') optomechanical systems. We calculate the linear optical properties    
of a high finesse cavity containing a thin dielectric membrane. We focus on the cavity's        
transmission, reflection, and finesse as a function of the membrane's position along the        
cavity axis and as a function of its optical loss. We compare these calculations with           
measurements and find excellent agreement in cavities with empty-cavity finesses in the range $10^{4}$ - $10^{5}$. The imaginary part of the membrane's index of refraction is found to      
be $\sim 10^{-4}$. We calculate the laser cooling performance of this system, with a            
particular focus on the less-intuitive regime in which photons ``tunnel'' through the membrane    
on a time scale comparable to the membrane's period of oscillation. Lastly, we present          
calculations of quantum non-demolition measurements of the membrane's phonon number in the low signal-to-noise regime where the phonon lifetime is comparable to the QND readout time.    
\end{abstract}

%\keywords{}

\maketitle

\section{Introduction}

Nearly all the optomechanical systems which have been studied to date consist of an optical cavity whose detuning is proportional to the displacement of some mechanical degree of freedom. The mechanical degree of freedom is most commonly the position of the end mirror of a Fabry-Perot cavity \cite{Braginsky1967, Caves1980, Bose1999, Marshall2003, McClelland2004, Metzger:2004, Arcizet2006, Zeilinger2006, Corbitt2007, Tittonen1999, Fabre1994, Vitali2004, Meystre1983, Gozzini1985} or the elongation of a waveguide  \cite{CarmonPRL2005, Wang2007, Kippenberg2007arXiv, Lehnert2008arXiv}. In these systems the radiation pressure has a physically intuitive form: it is a force which acts on the mechanical degree of freedom and is proportional to the instantaneous intracavity optical power.

Recently a new type of optomechanical system has been described in which the mechanical degree of freedom is a flexible, partially transparent object (such as a dielectric membrane) placed inside a Fabry-Perot cavity \cite{Harris08, Karrai2007, MeystreJOSAB:1985, Meystre2007arXiv, Meystre2008arXiv, Meystre2008arXiv_2}. In this type of system the cavity detuning (and hence the radiation pressure) is periodic in the membrane displacement. 

Here we analyze several aspects of such a ``dispersive'' optomechanical device. We calculate its linear optical properties (transmission, reflection, and finesse) as a function of experimentally relevant parameters, and compare these calculations with experiments. We demonstrate a dispersive optomechanical device with a finesse $F = 150,000$, and argue that it should be possible to realize $F = 500,000$ using present-day technology.

We also present calculations of the radiation pressure-induced cooling and heating in these  systems. Because dispersive optomechanical systems consist of a compound optical cavity, their laser cooling is more complicated than in the more familiar ``reflective'' optomechanical devices described, e.g., in refs \cite{Braginsky1967, Caves1980, Bose1999, Marshall2003, McClelland2004, Metzger:2004, Arcizet2006, Zeilinger2006, Corbitt2007, Tittonen1999, Fabre1994, Vitali2004, Meystre1983, Gozzini1985, CarmonPRL2005, Wang2007, Kippenberg2007arXiv, Lehnert2008arXiv}. Lastly, we consider phonon quantum non-demolition (QND) measurements in dispersive optomechanical systems \cite{Harris08, Santamore2004}. We focus in particular on phonon QND measurements with low signal-to-noise ratios, and consider how quantum effects might be manifest in such non-ideal experiments.
 
\section{Linear optical properties: calculations}
\begin{figure}
\begin{center}
\includegraphics[width=4in]{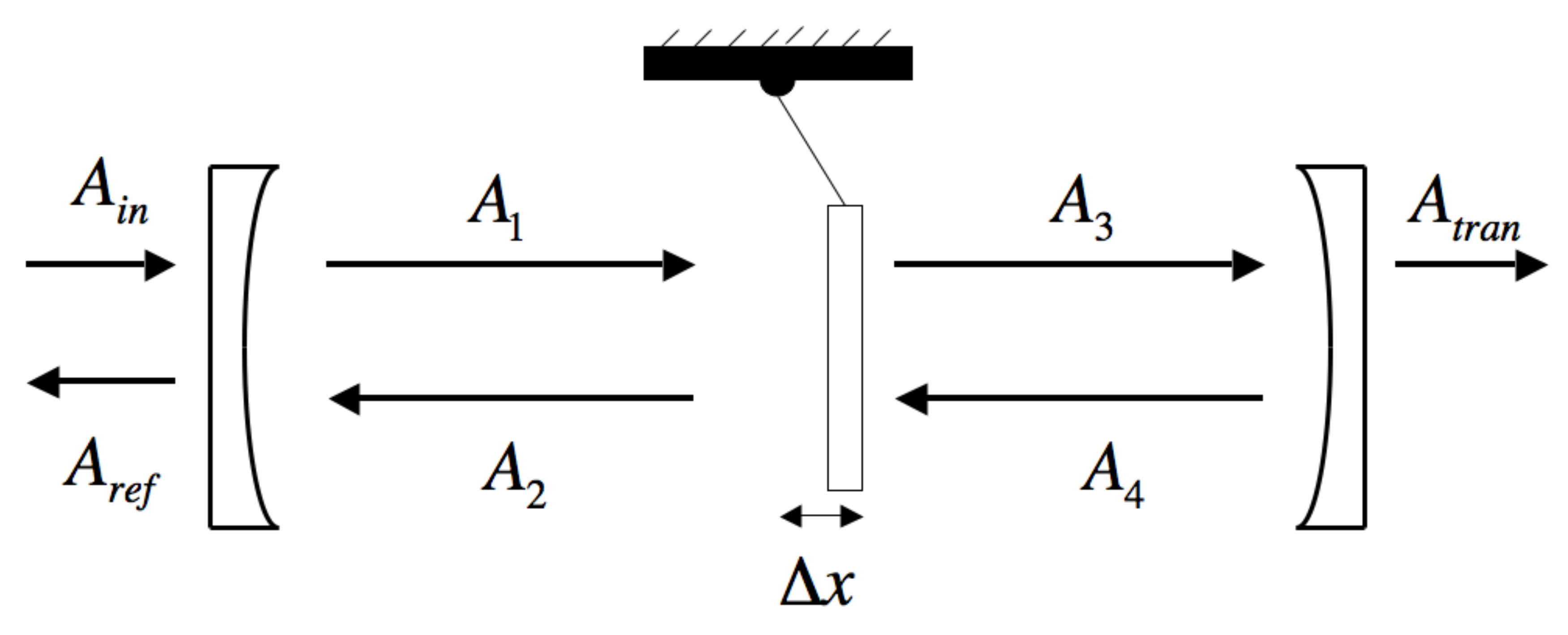}
\caption{Schematic illustration of the dispersive optomechanical system. The membrane is represented as a mass on a pendulum.  Enclosing the membrane are the end mirrors that define a high finesse cavity.  $A_{in}$, $A_{ref}$, $A_{tran}$, $A_{1}, A_{2}, A_{3}, A_{4}$ are the incident, reflected, transmitted and circulating fields. \label{cavity}}
\end{center}
\end{figure}

The geometry of the dispersive optomechanical devices considered in this paper is shown in figure \ref{cavity}. Our one-dimensional model consists of two cavity end mirrors with electric field reflectivity $r$ and transmission $t$ (the two cavity mirrors are assumed identical in this paper, but the extension to unequal  mirrors is straightforward). The dielectric membrane placed between the two end mirrors has a thickness $L_{d}$ and index of refraction $n$. The membrane's electric field reflectivity $r_{d}$ and transmission $t_{d}$ are then given by \cite{Brooker}

\begin{equation}
r_{d} = \frac{(n^{2} -1)\sin k n L_{d}}{2 i n \cos k n L_{d} + (n^{2} +1)\sin k n L_{d}}
\label{reflection_amplitude}
\end{equation}

\begin{equation}
t_{d }= \frac{2 n}{2 i n \cos k n L_{d} + (n^{2} +1)\sin k n L_{d}}
\label{transmission_amplitude}
\end{equation}
where $k$ is the wavenumber of the light incident on the membrane.

Note that $r_{d}$ and $t_{d}$ are in general complex (reflecting the phase shift acquired by light reflected from or transmitted through a dielectric slab). If $n$ is real, then $\mid r_{d} \mid ^{2} + \mid t_{d} \mid ^{2} = 1$. However in general $n$ will be complex, with the imaginary part determining the membrane's optical absorption.

To find the transmission and reflectivity of the cavity as a whole, we solve the following system of equations:
\numparts
\begin{equation}
	A_{1} =  
		itA_{in}+rA_{2}e^{ikL_{1}} 
	\label{field1} 
\end{equation}
\begin{equation}
	A_{2}  = 
r_{d}A_{1}e^{ikL_{1}}+it_{d}A_{4}e^{ikL_{2}} \\
\end{equation}
\begin{equation}
	A_{3}  =  
		it_{d}A_{1}e^{ikL_{1}}+r_{d}A_{4}e^{ikL_{2}} \\
\end{equation}
\begin{equation}
	A_{4}  =  
		rA_{3}e^{ikL_{2}} \\
\end{equation}
\begin{equation}
	A_{refl}  = 
		itA_{2}e^{ikL_{1}}+rA_{in} \\
\end{equation}
\begin{equation}
	A_{tran}  =  
		itA_{3}e^{ikL_{2}} 
		\label{fieldTran} \\
\end{equation}
\endnumparts
Here $A_{1}$ through $A_{4}$ are the electric field amplitudes of travelling waves in the cavity (as shown in figure 1), and $A_{in}$, $A_{refl}$, and $A_{trans}$ are the amplitudes of the incident, reflected, and transmitted waves. $L_{1}$ and $L_{2}$ are the lengths of the left- and right-hand halves of the cavity shown in Figure \ref{cavity} \cite{Dandliker1969}.

Since we are primarily interested in cases where the cavity finesse is high and the membrane absorption is low, we find the cavity resonance frequencies by solving for the eigenfrequencies of the closed lossless cavity (i.e., assuming $r = 1$ and Im$(n) = 0$). The solution gives:

\begin{equation}
\delta_{T}^{(0)} = 2 \phi_{r}+2 \cos^{-1}(\mid r_{d} \mid \cos \delta)
\label{dispersion_equation}
\end{equation}
where $\delta_{T}^{(0)}$ is the cavity's resonance frequency scaled by $2 \pi / f_{FSR}$ and $f_{FSR}$ is the cavity's free spectral range. The scaled membrane position is $\delta \equiv 2k\Delta x$.   $\phi_{r}$ is the complex phase of $r_{d}$.  From this expression it is clear that the magnitude of the membrane's reflectivity $\mid r_{d} \mid$ determines the dependence of $\delta_{T}^{(0)}$ on the membrane position. $\phi_{r}$, the complex phase of $r_{d}$, sets an overall offset to $\delta_{T}^{(0)}$.  

\begin{figure}
\begin{center}
{\scalebox{.6}{\includegraphics{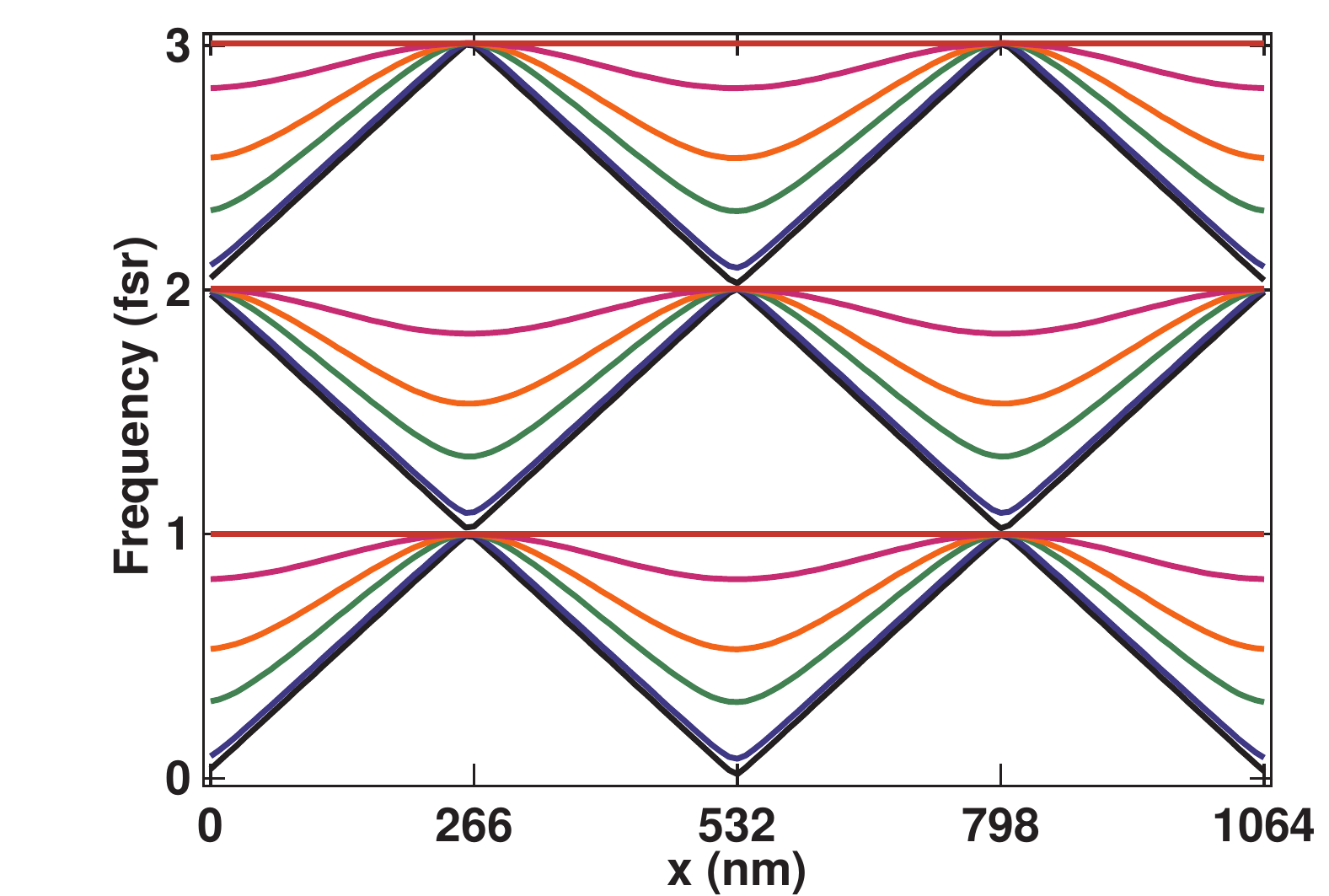}}}
\caption{Cavity detuning as a function of membrane position. The resonant frequencies of three of the cavity's longitudinal modes are plotted (in units of the cavity free spectral range) for several values of the membrane reflectivity.  The membrane's power reflectivities $\mid r_{d} \mid^{2}$ are the following: red: 0.000, orange: 0.080, yellow: 0.450, green: 0.773, blue: 0.982, and black: 0.999. For this calculation the various reflectivites were realized by fixing the membrane thickness and varying the index of refraction.
\label{theory_band}}
\end{center}
\end{figure}

The cavity detuning versus membrane position is shown for several membrane reflectivities in figure \ref{theory_band}. For illustrative purposes the membrane reflectivity was varied by setting the membrane thickness to be unphysically small ($L_d = 0.01$ nm) and varying $n$.

The analytic expressions for the transmission through the cavity and reflection from the cavity are also straightforward, but are too cumbersome to display here. We have not found a simple expression for the cavity finesse; instead we estimate it numerically from the linewidth of the transmission resonances. 

\begin{figure}
\begin{center}
$\begin{array}{c}
{\scalebox{.8}{\includegraphics{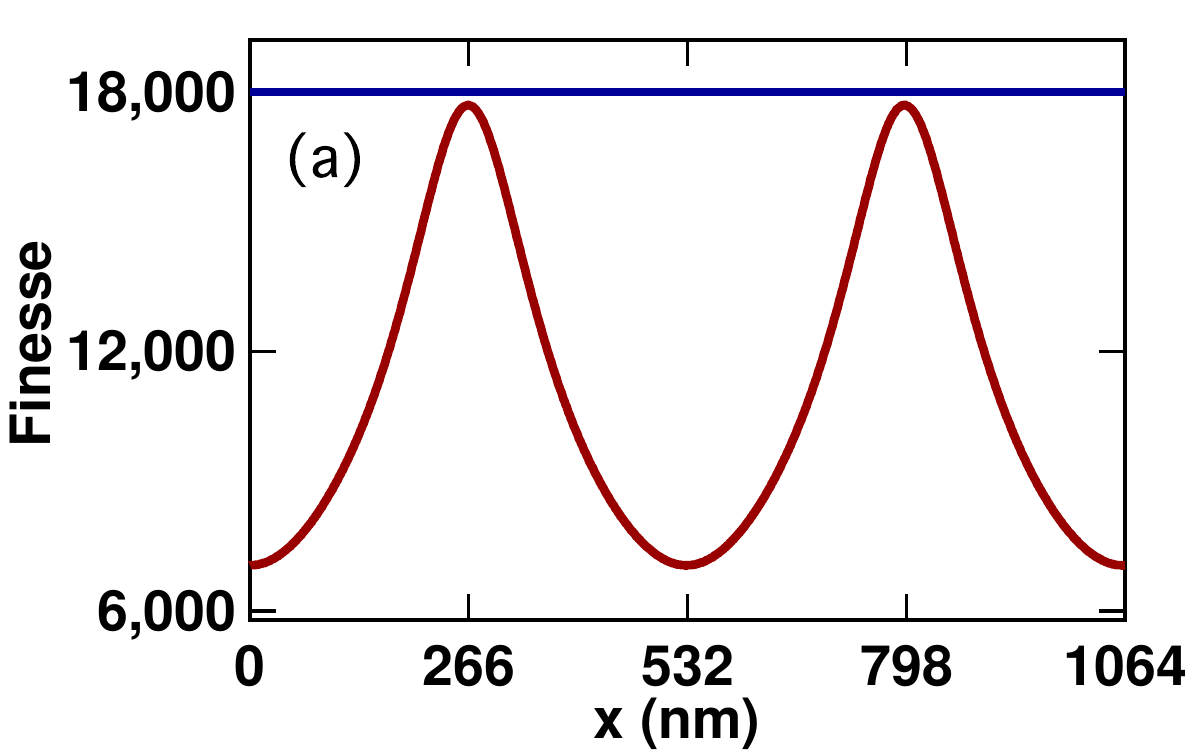}}} \\
%{\mathbf(a)} \\ 
{\scalebox{.8}{\includegraphics{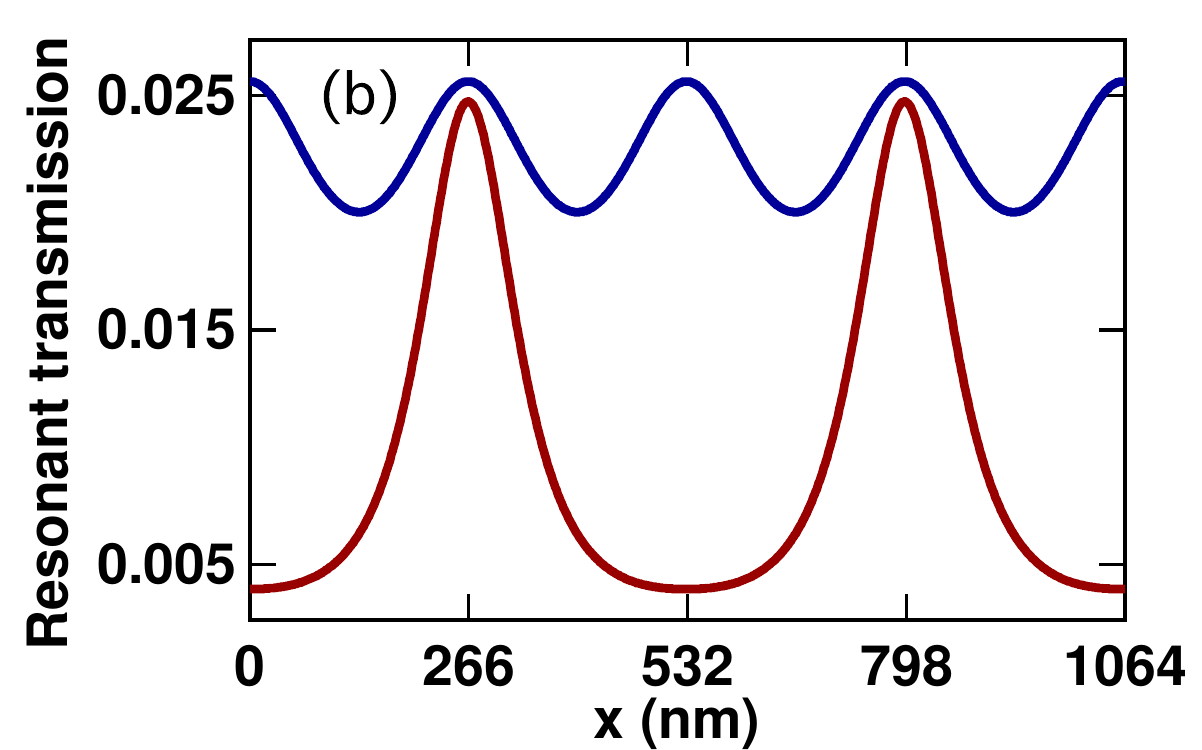}}} \\ 
%{\mathbf(b)}\\
{\scalebox{.8}{\includegraphics{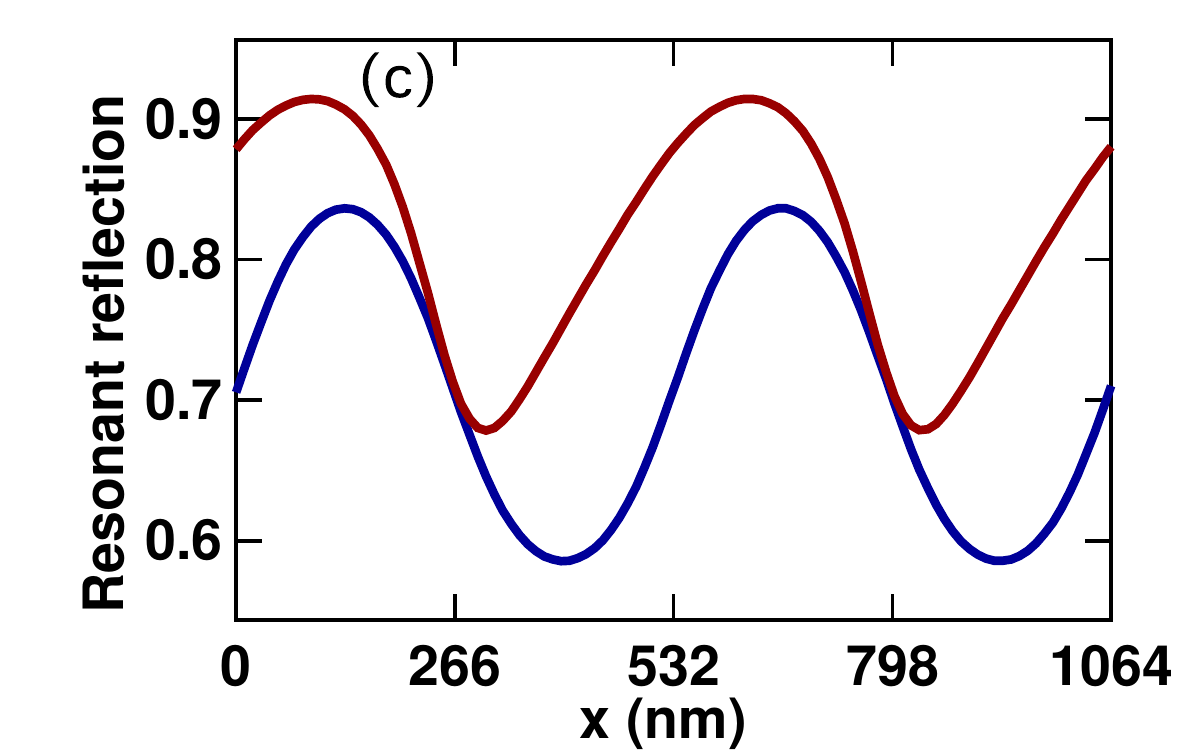}}} \\
%{\mathbf(c)} \\
% {\scalebox{.5}{\includegraphics{../igor/fin_theory}}} \\ 
%{\mathbf(a)} \\
%{\scalebox{.5}{\includegraphics{../igor/tran_theory}}} \\
%{\mathbf(b)}\\
%{\scalebox{.5}{\includegraphics{../igor/ref_theory}}} \\
%{\mathbf(c)}\\
\end{array}$
\end{center}
\caption{\label{theoretical} Calculated finesse (a), resonant transmission (b), and resonant reflection (c) of the dispersive optomechanical cavity as a function of membrane position.  The blue curves correspond to a lossless membrane, while the red curves correspond to a lossy membrane. Note that all the calculations assume the same (non-zero) loss in the end mirrors. This leads to a reflection and transmission which do not add to unity, and which depend differently upon the membrane position. The parameters used in this calculation are given in the text.}
\end{figure}

Figure \ref{theoretical} shows the finesse, resonant transmission (i.e., the transmission when the laser is resonant with the cavity), and the resonant reflection as a function of membrane position. These plots assume $L_{d} = 50$ nm, $\frac{2\pi}{k}$ = 1064 nm, $r = 0.99991$ and $t = 5.28 \times 10^{-3}$ (i.e., the power transmission of each end mirror is 16$\%$ of what it would be if it were lossless) corresponding to an empty-cavity finesse of $18,000$. In each plot the blue curve corresponds to $n = 2.2$ (i.e., a lossless membrane), while the red curve corresponds to $n = 2.2 + 1.5 \times 10^{-4} i$ (i.e., a membrane with some optical loss).

The  differences between the curves for the lossless membrane (blue) and the lossy membrane (red) can be understood qualitatively. Placing a lossless membrane inside a cavity does not alter the rate at which photons leak out of the cavity; as a result the blue curve in figure \ref{theoretical}(a) is flat. However the position of a lossless membrane does modulate the resonant transmission (figure \ref{theoretical}(b)) because the cavity eigenmodes will be modified by the membrane. Cavity modes primarily localized on the right-hand side of the cavity in figure \ref{cavity} will leak primarily out of the right-hand mirror, leading to an increased transmission coefficient for the cavity as a whole. Likewise, modes localized predominantly on the left-hand side of the cavity will couple primarily to external modes to the left of the cavity, leading to an increased reflection coefficient (figure \ref{theoretical}(c)).

When $n$ is complex, intracavity photons can be lost to the membrane absorption. This additional loss process lowers the cavity finesse by an amount proportional to the overlap of the cavity mode with the membrane, giving rise to the dips in the red curve of figure \ref{theoretical}(a) when the membrane is positioned at an antinode of the cavity mode. Note that for the parameters used in this calculation, the finesse is not appreciably reduced from its empty-cavity value if the membrane is placed at a node of the optical field (i.e., corresponding to the peaks in the red curve of figure \ref{theoretical}(a).

The loss of photons due to absorption in the membrane also prevents photons from transiting the cavity. As a result the resonant transmission has pronounced dips when the membrane is at an antinode (red curve in figure \ref{theoretical}(b)). The reflection signal (figure \ref{theoretical}(c) arises from interference between intracavity light leaking out through the left-hand end mirror (which is affected by the membrane's absorption) and light promptly reflected from the left-hand end mirror (which is not). As a result its form is less intuitive, with membrane absorption leading to an asymmetric dependence on membrane position.

\section{Linear optical properties: measurements}
\subsection{Experimental Setup}

\begin{figure}
\begin{center}
\scalebox{.5}{\includegraphics{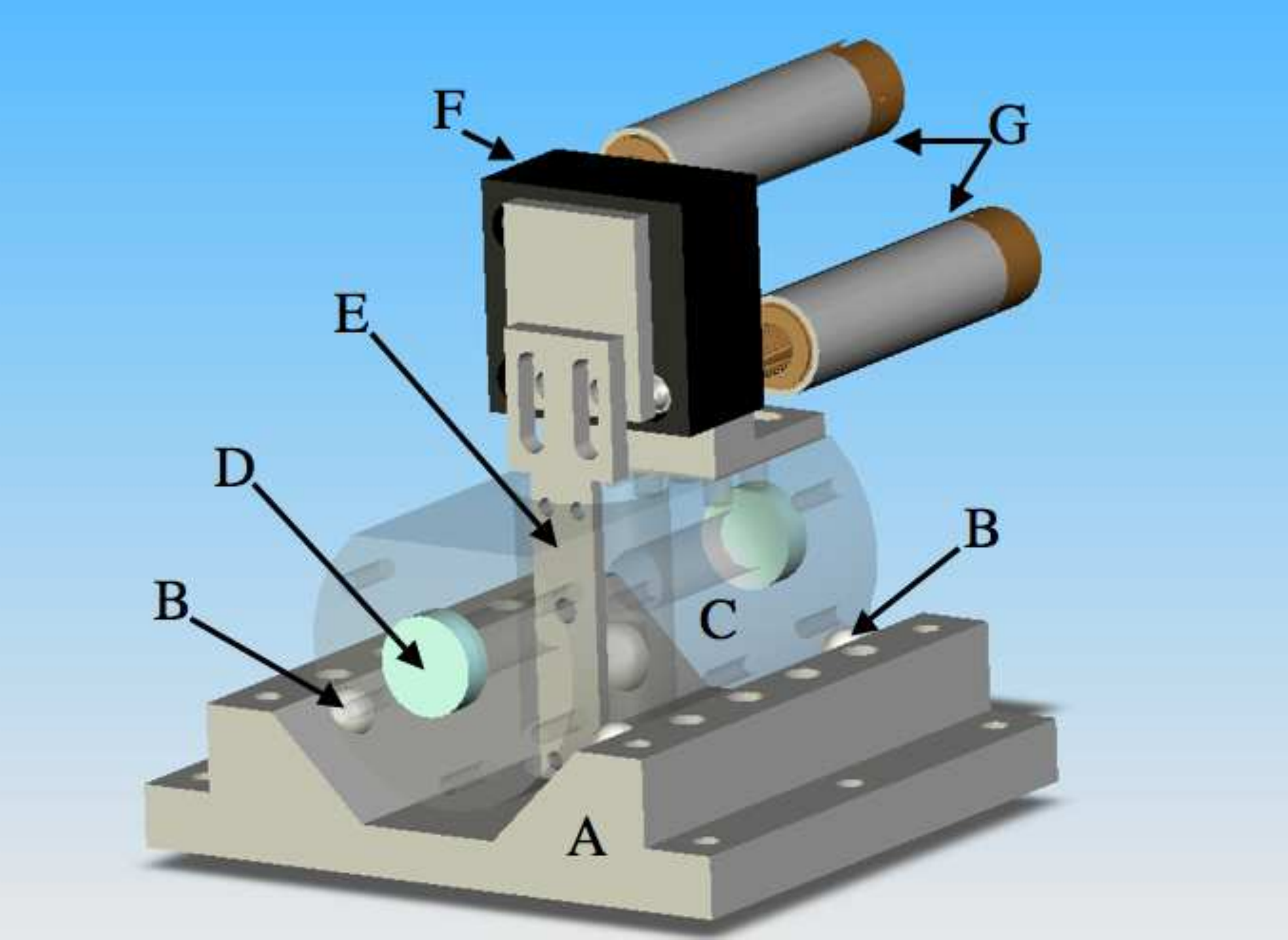}}
\caption{Illustration of the optical cavity. The Invar cavity support (A) is mounted to the inside     
of the vacuum chamber. A series of alumina spheres (B) are mounted in cone-shaped            
recesses to provide kinematic mounting between the support and the Invar cavity spacer     
(C). The end mirrors (D) define the optical cavity. The membrane and piezoelectric           
elements are mounted to the Invar arm (E) which is in turn mounted to the tilt stage (F).    
The tilt stage can be adjusted \textit{in situ} by two motorized actuators (G).
\label{assembly}}
\end{center}
\end{figure}

\begin{figure}
\begin{center}
\includegraphics[width=3.5in, height = 4in]{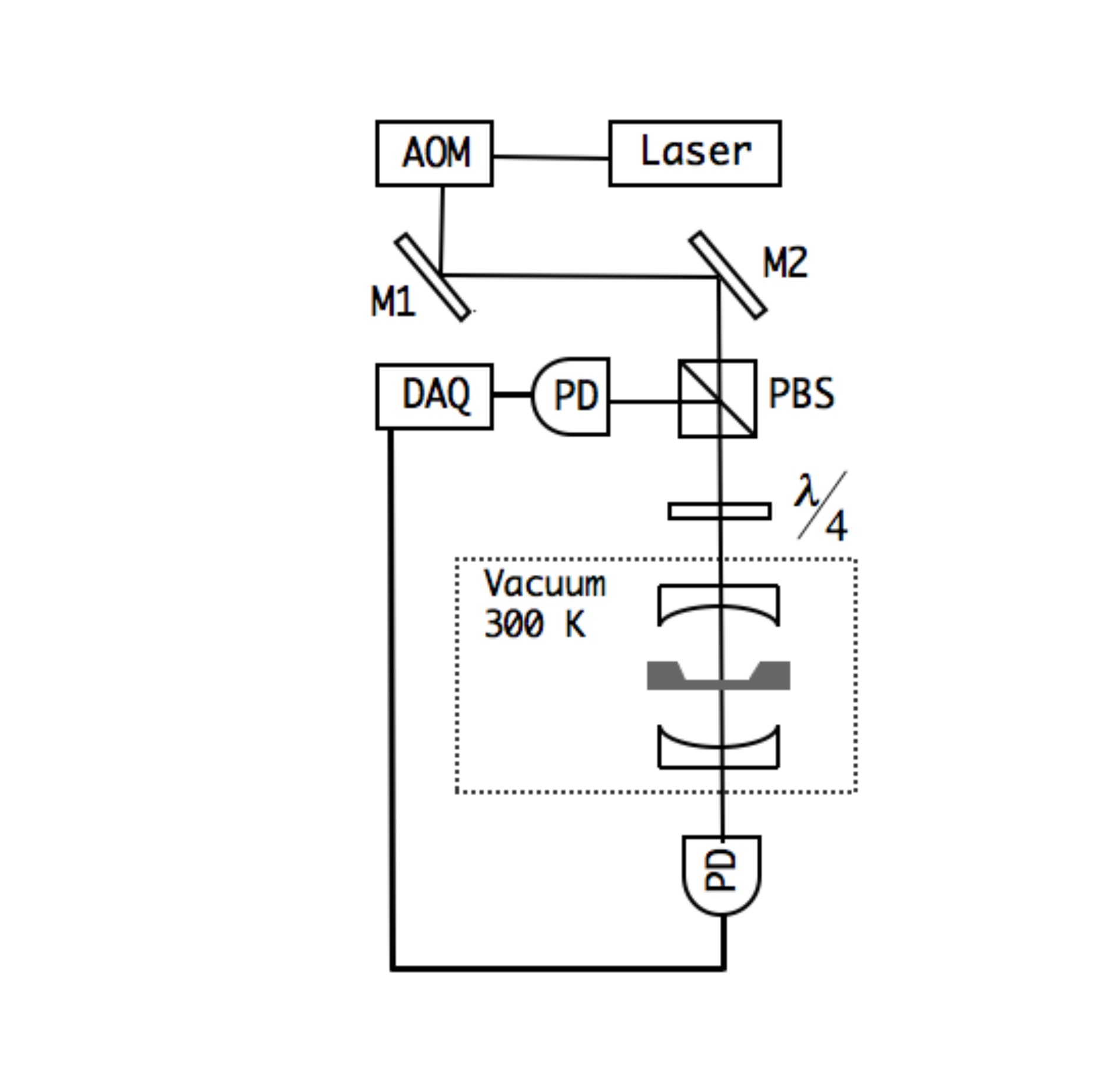}
\caption{A schematic of the experimental setup.  The membrane is depicted in grey between the two high finesse cavity end mirrors. The transmission and reflection are monitored with photodetectors (PD) and the signals are sent to a data acquisition (DAQ) card. 
\label{schematic}}
\end{center}
\end{figure}

Schematic illustrations of our experiment are shown in figures \ref{assembly} and \ref{schematic}. Laser light is produced by a Nd:YAG laser (Innolight, Hannover, Germany) with wavelength $\lambda = 1064$ nm. The light passes through an acousto-optic modulator (AOM), and the first-order beam is sent to the optomechanical cavity via the steering mirrors M1 and M2. The cavity is formed by two dielectric mirrors each with a 5 cm radius of curvature (coated by Advanced Thin Films, Longmont, CO, USA). The mirrors are mounted to a cylindrical Invar spacer 6.7 cm long with a hole drilled along its axis to accomodate the cavity mode.

The Invar spacer has a second hole drilled perpendicular to the cavity axis, allowing us to introduce the dielectric membrane into the waist of the cavity mode. The membrane is mounted on two piezoelectric elements. The first allows us to apply high frequency ($\simeq 300$ kHz), small amplitude ($\sim 1$ nm) oscillations to the membrane (e.g., to excite its mechanical eigenmodes). The second piezo allows us to translate the membrane by roughly 2 $\mu$m along the cavity axis. The piezo elements are in turn mounted to a tilt stage (Thorlabs, Newton, NJ, USA KM05) which is rigidly attached to the Invar spacer. The tilt stage includes vacuum-compatible motorized actuators (Thorlabs, Newton, NJ, USA Z612V) allowing us to adjust \textit{in situ} the angular alignment of the membrane relative to the cavity axis. In practice we have found that the membrane needs to be aligned to roughly 5 arcseconds in order to achieve the highest finesse described below.

The membrane used in these experiments is a commercial, 50 nm thick, 1 mm $\times$ 1 mm SiN x-ray window (Norcada, Edmonton, AB, Canada). The membrane is supported by a 200 $\mu$m thick Si frame. The exceptional mechanical properties of these membranes have been described elsewhere \cite{zwickl2008}.

In practice we first align the cavity with the membrane removed. Then the membrane is inserted and its tilt and transverse position are adjusted until good transmission through the cavity is achieved.

For most of the measurements presented here, the cavity is mounted inside a vacuum chamber which is pumped down to $\sim 10^{-6}$ torr by an ion pump. Good vacuum is crucial to maintaining the membrane's high mechanical quality factor.

\subsection{Measurements}

We monitor the optical power reflected from and transmitted through the cavity using the photodiodes shown in figure \ref{schematic}. We can also determine the transverse profile of the cavity mode by imaging the transmitted beam with a video camera.  

\begin{figure}
\begin{center}
$\begin{array}{cc}
{\scalebox{.6}{\includegraphics{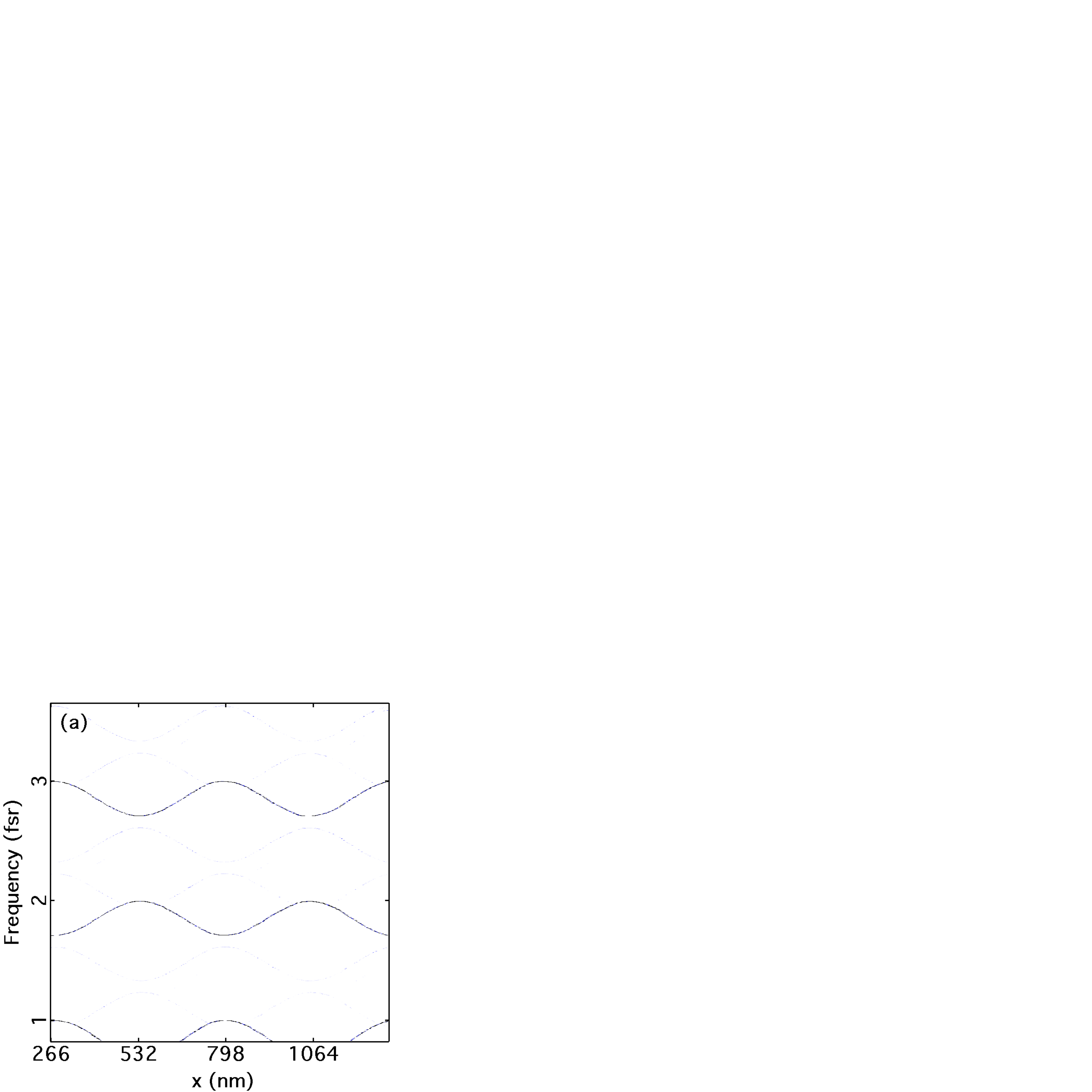}}} &  \scalebox{.6}{\includegraphics{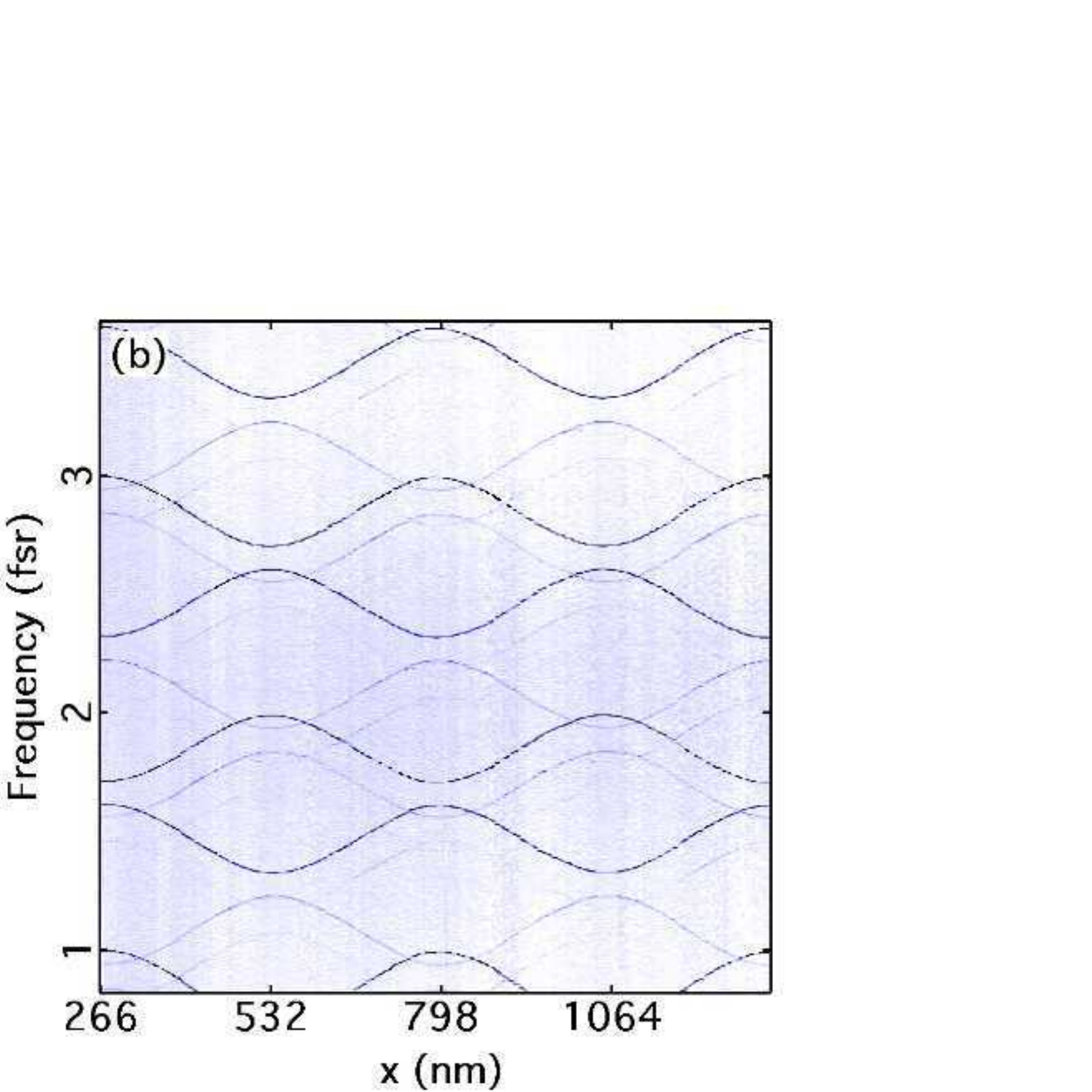}} \\
%{\mathbf(a)} & {\mathbf(b)}  \\
\end{array}
\newline
\begin{array}{c}
\scalebox{.6}{\includegraphics{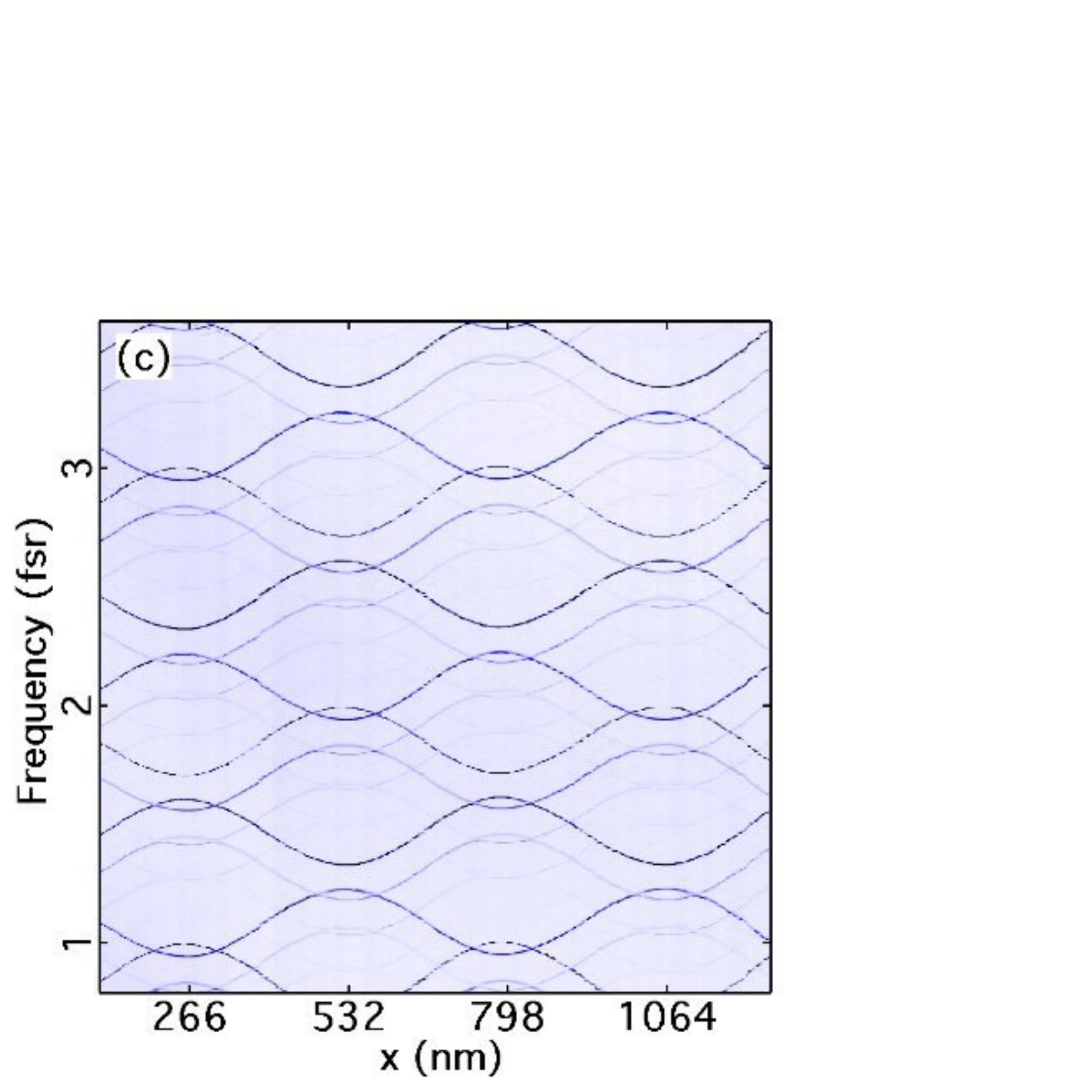}} \\
%{\mathbf(c)}  \\
\end{array}$
\end{center}
\caption{Measured transmission (on a linear scale) as a function of laser frequency (in units of the cavity free spectral range) and the membrane position. In (a) the input beam is coupled almost exclusively to the cavity's TEM$_{00}$ (transverse Gaussian) mode.  In (b) the input beam is coupled to both the TEM$_{00}$ and the TEM$_{m+n=1}$ transverse modes. In (c) the input beam is coupled to several transverse modes, including the Gaussian.\label{dispersion}}
\end{figure}

Figure \ref{dispersion}(a) shows the optical power transmitted through the cavity as a function of the laser frequency and the membrane position when the laser is mode-matched to the cavity's TEM$_{0,0}$ modes. The dark bands (indicating high transmission) correspond to the cavity's resonant frequencies. Comparison of these resonant bands with (\ref{dispersion_equation}) gives a value of $\mid r_c \mid = 0.35$.  As with all the data in this paper, the calibration of the membrane position is taken from the assumption that the features in the data are periodic in the membrane displacement with period $\pi/k$.

Figure \ref{dispersion}(b) shows similar data, but taken with the input beam aligned in such a way as to couple to the TEM$_{0,0}$ mode and the nominally degenerate doublet consisting of the TEM$_{0,1}$ and TEM$_{1,0}$ modes \cite{Anderson1984}. In figure \ref{dispersion}(c), the input is realigned to couple into still more of the cavity's modes. The cavity spectra shown in these figures can be easily explained using the standard description of higher-order transverse modes in optical cavities \cite{Lasers_Siegman}, and a detailed description will be given in a future publication.

\begin{figure}
\begin{center}
$\begin{array}{c}
{\includegraphics[width=3.5in]{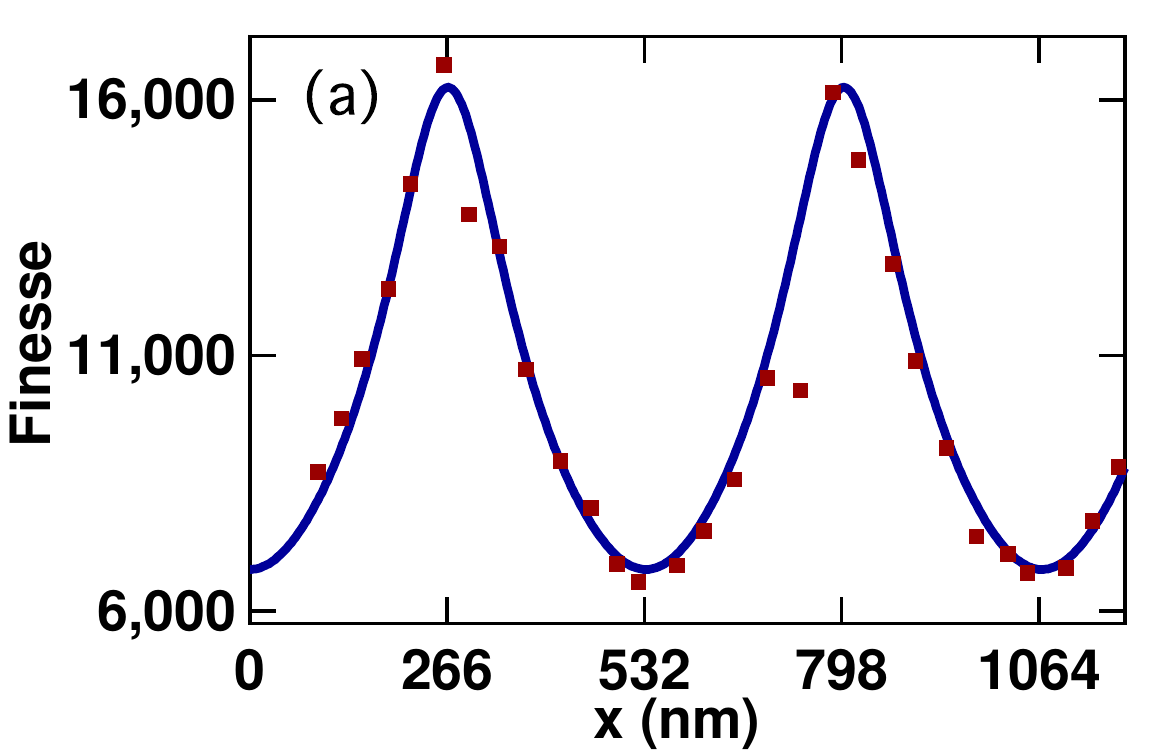}} \\
%{\mathbf(a)}\\
{\includegraphics[width=3.5in]{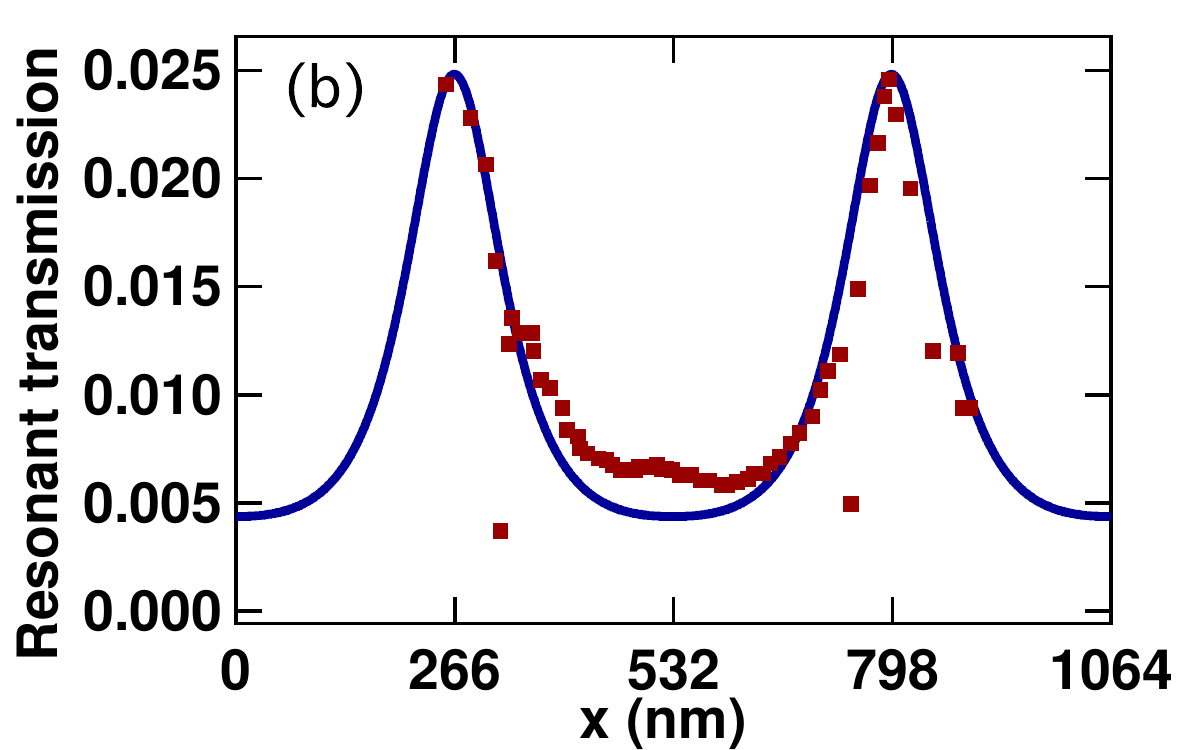}} \\
%{\mathbf(b)}\\
{\includegraphics[width=3.5in]{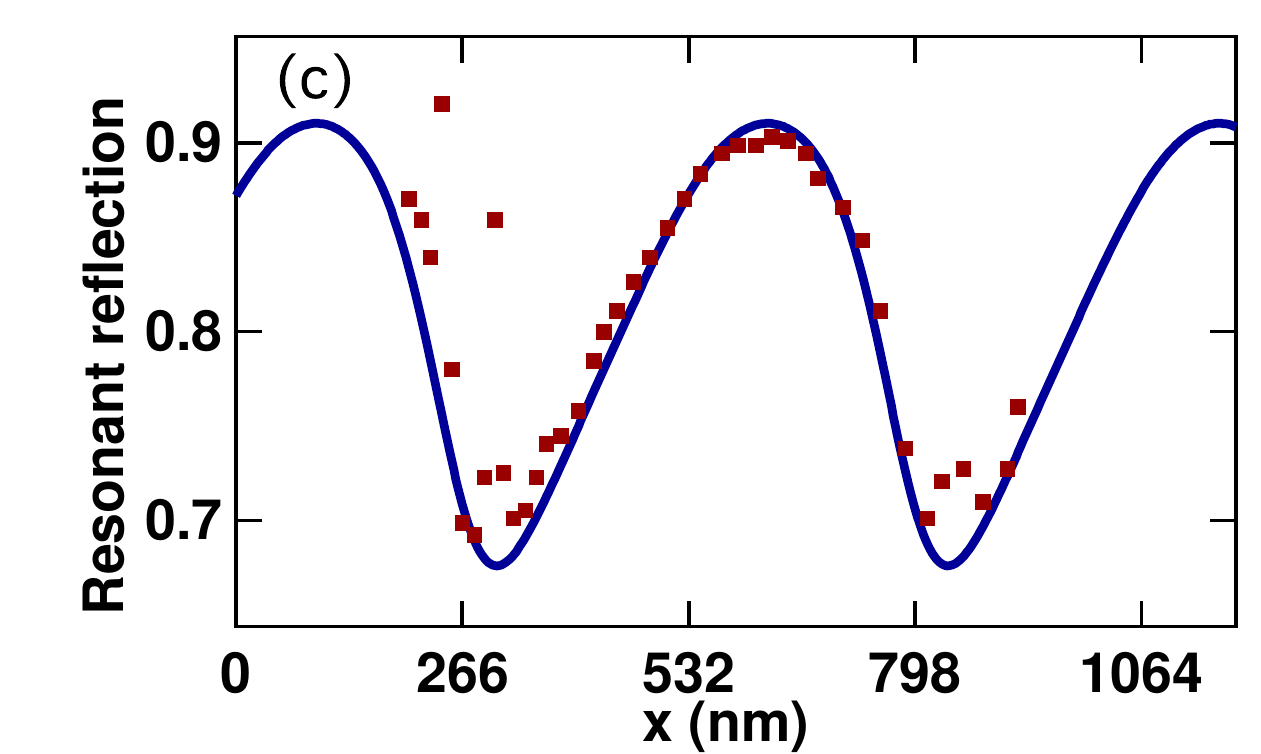}} \\
%{\mathbf(c)}\\
\end{array}$
\end{center}
\caption{\label{measured} Measurements and fits of the cavity's finesse (a), resonant transmission (b), and resonant reflection (c) as a function of membrane position.  The solid curves are fits to the data.  These fits assume an empty-cavity finesse of 16,500 and give an imaginary part of the membrane's index of refraction of $1.5 \times 10^{-4}$. }
\end{figure}

The cavity finesse is determined from cavity ringdown measurements. In these, the laser frequency is swept slowly while the optical power transmitted through the cavity is monitored. When the transmitted signal exceeds a pre-determined threshold (indicating that the laser is coming into resonance with the cavity)  the AOM switches off the input beam and the transient leakage of light out of the cavity is monitored. This decay has a single exponential form \cite{Harris08, zwickl2008} whose time constant, $\tau$, is related to the cavity finesse, $F$, via $F=2 \pi f_{FSR} \tau$. Figure  \ref{measured}(a) shows the finesse of the cavity's TEM$_{0,0}$ mode as a function of membrane position. The solid line in figure \ref{measured}(a) is a fit to the calculation described above and shown in figure \ref{theoretical}(a).

Figures \ref{measured}(b) and (c) show the cavity transmission and reflection on the TEM$_{0,0}$ resonance as a function of the membrane position. The solid lines are fits to the data using the calculations described above. The data in Figure 7 were taken with the device in vacuum.

The fits in figure \ref{measured} assume $L_d = 50$ nm, that the empty cavity finesse is 16,500, and that the end mirrors' transmission and reflection coefficients are  $t = 5.52 \times 10^{-3}$, and $r = 0.99991$ (consistent with measurements of the cavity when the membrane was removed). Given these constraints, the fits yield $n = 2.2 + 1.5 \times 10^{-4} i$. The agreement between the data and fits indicates that our simple model does a reasonable job of describing the system. The few anomalous data points in figures \ref{measured}(a), (b) and (c) correspond to membrane positions in which the TEM$_{0,0}$ mode becomes degenerate with other cavity modes (see, e.g., figure \ref{dispersion}(c)). Such degeneracies are not accounted for in our simple one-dimensional model.

\subsection{Discussion}

\begin{figure}
\begin{center}
\includegraphics[width=3.5in]{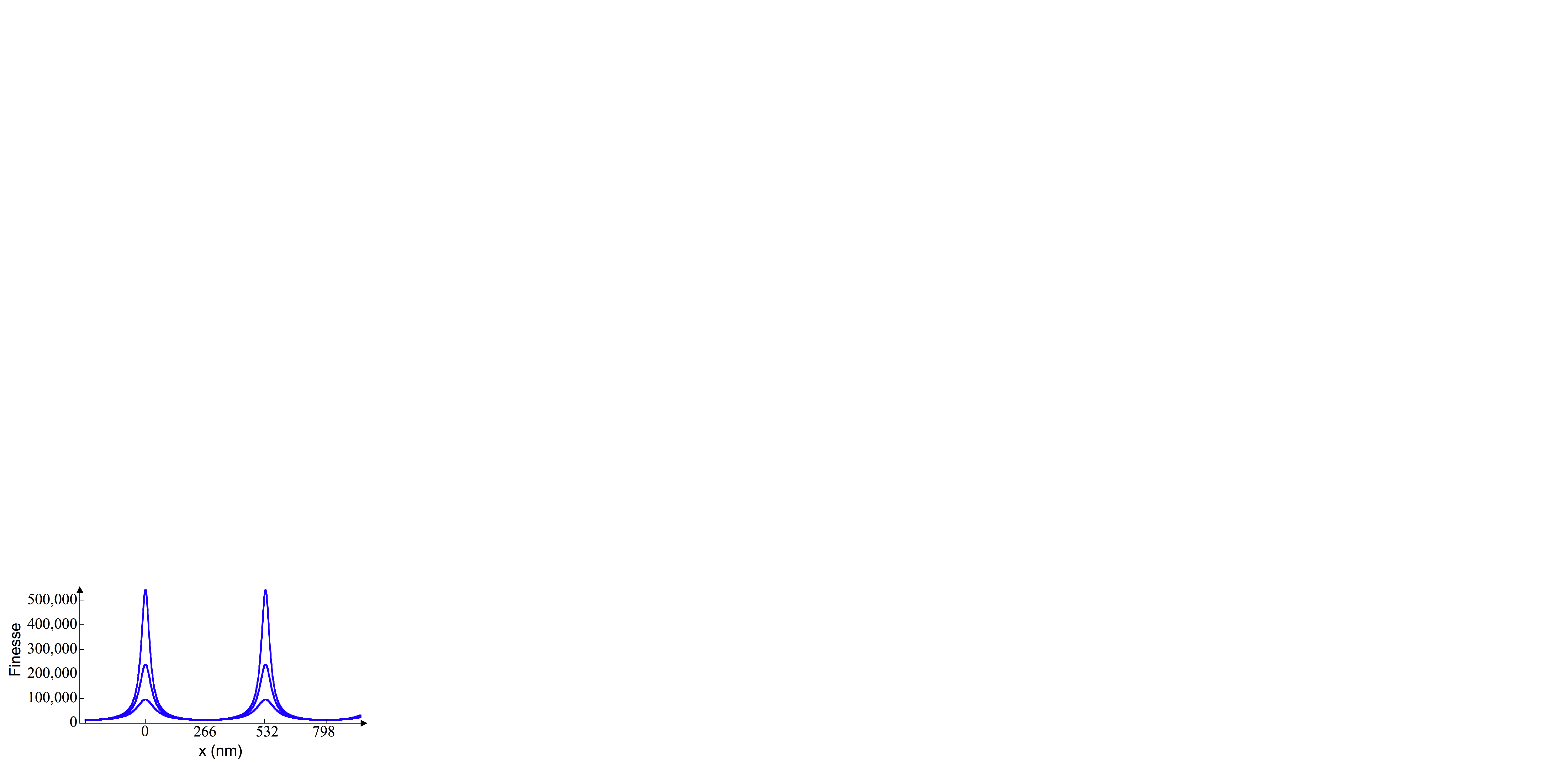}
\caption{A plot of the calculated finesse as a function of membrane position for different    
cavity end mirrors. All three curves assume the membrane's index of refraction is $n = 2.15     
+ 1.5 \times 10^{-4} i$. The empty cavity finesse is taken to be 100,000 (lowest curve), 314,000       
(middle curve), and 1,000,000 (uppermost curve). When the membrane is positioned at a node,     
the device's finesse can be $> 500,000$.
\label{finesse limit}}
\end{center}
\end{figure}

The loss in the membrane places limits on the maximum obtainable finesse. However the data in figure \ref{measured}(a) indicates that this limit depends strongly upon where the membrane is placed relative to the cavity nodes and antinodes. In order to determine this limit quantitatively, we use the value of Im$(n) = (1.5 \pm 0.1) \times 10^{-4}$ extracted from the data and fits in figure \ref{measured}(a) to calculate the finesse of hypothetical devices which are identical to the ones measured here but with higher reflectivity end mirrors. The result is shown in figure \ref{finesse limit}, and indicates that for experiments in which the membrane can be placed at a node, it should be possible to achieve $F \simeq 500,000$ with state-of-the-art end mirrors (i.e., those corresponding to an empty-cavity finesse of $1,000,000$). For experiments in which the membrane must be placed away from a node, figure \ref{measured} (a) indicates that the optical loss in these membranes cannot be compensated for by better end mirrors.

\begin{figure}
\begin{center}
\scalebox{1}{\includegraphics{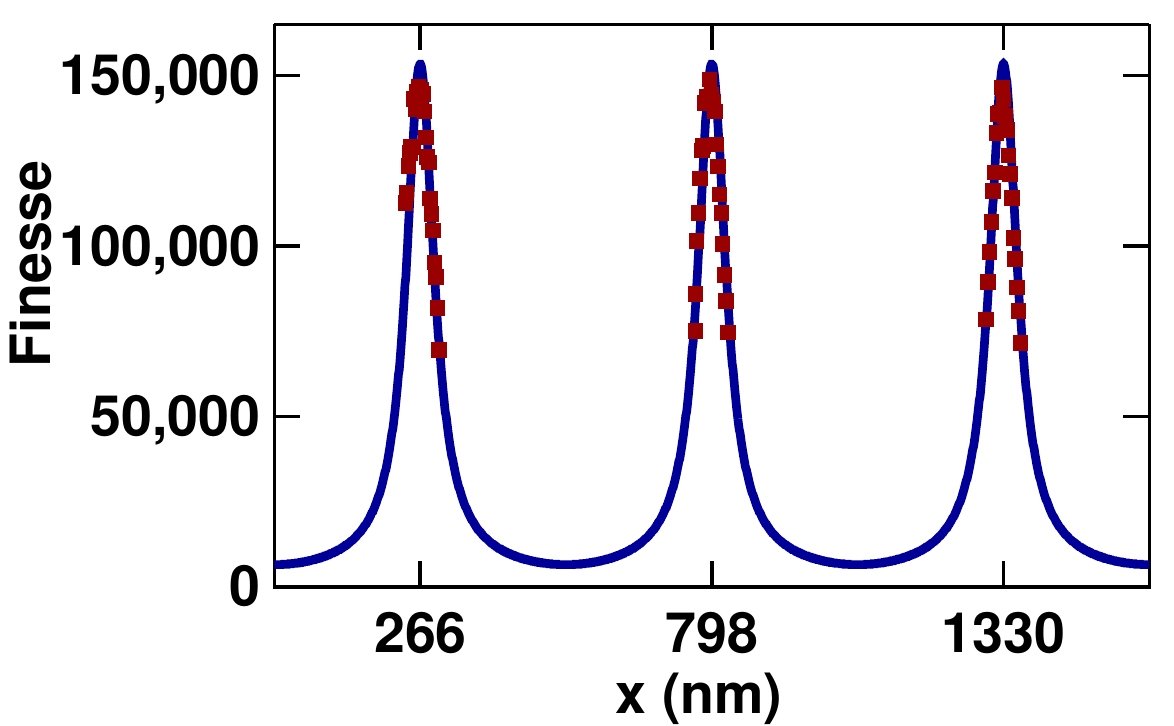}}
\caption{Finesse as a function of membrane position in a higher-finesse cavity. Similar to    
figure \ref{theoretical} (a), but taken with end mirrors giving an empty cavity finesse $\approx 205,000$. The fit assumes an empty cavity finesse of 205,000 and gives Im$(n) = 2.3 \times 10^{-4}$.
\label{hiFinesse}}
\end{center}
\end{figure}

To test the prediction shown in figure \ref{finesse limit} we replaced the end mirrors used in the measurements described above with end mirrors giving a measured empty-cavity finesse $205,000 \pm 10,000$. Figure \ref{hiFinesse} shows the resulting finesse (measured in air) as a function of membrane position, along with a fit which gives Im$(n) = (2.3 \pm 0.06) \times 10^{-4} $  (the empty cavity finesse is set to 205,000). Although these fits indicate that there may be some sample-to-sample variation in the membranes' absorption, the overall level of agreement between the four sets of data (in figures \ref{measured} and \ref{hiFinesse}) and the theory indicates that extrapolation to still higher-reflectivity end mirrors is justified, and that it should be possible to realize a dispersive optomechanical system with a cavity finesse $\approx 500,000$ (see figure \ref{finesse limit}). We have assumed $L_{d} = 50$ nm throughout this discussion; still higher finesse could be achieved with thinner membranes, albeit with a decrease in optomechanical coupling.

This high finesse is realized when the membrane is positioned at a node of the intracavity field. This arrangement is ideal for the phonon QND measurements described in \cite{Harris08} and below. However it also corresponds to a point at which the adiabatic radiation pressure is identically zero, and so may seem to preclude the realization of more familiar optomechanical effects such as laser cooling. In fact the situation is somewhat more complicated, as described in the following section.

%Florian's contribution.
\section{Optomechanical cooling and heating}

In this section, we obtain the optomechanical cooling and heating rates
 (as well as the optomechanical spring effect) in a membrane-in-the-middle
 (MIM) setup.
 We do so by solving the linearized coupled classical equations of
 motion for the membrane coordinate  $x$ and the amplitudes $\alpha_{L,R}$ 
 of two optical modes in the left and the right halves of
 the cavity. This means we are assuming that the membrane reflectivity $\mid r_c \mid\rightarrow 1$ (corresponding, e.g. to the black curve in figure \ref{theory_band}, and that the left and right half cavities are nearly degenerate (corresponding to membrane positions near the avoided crossings in figure \ref{theory_band}). Although this regime has not been achieved experimentally, it is in this situation that the most striking deviations from the usual (i.e., ``reflective'') setup are to be expected and indeed are realized.
 Note that modeling a device in which $\mid r_c \mid \ll 1$ (corresponding to an approximately sinusoidal relationship between $\delta_{T}^{(0)}$ and $\Delta x$)  would require including the contributions from many different modes, not just two.

To set the scene, we first review the calculation of the linearized optomechanical dynamics for the simpler, well-understood case of a single cavity mode in a ``reflective'' optomechanical device \cite{2007_01_Marquardt_CantileverCooling, Kippenberg2007}. We will present it in a way that prepares us for the derivation involving
 the ``dispersive'' optomechanical device.

\subsection{Linearized dynamics of standard optomechanical systems}

We have the following equations of motion for the membrane coordinate $x$ and the complex light amplitude $\alpha$ (rescaled such that $\alpha=1$  at resonance, and taken in a frame rotating at the laser frequency  $\omega_{L}$):

 \begin{eqnarray}
\dot{\alpha} & = & i(\Delta-\omega'x)\alpha+\frac{\kappa}{2}(1-\alpha)\\
\ddot{x} & = & -\omega_{M}^{2}(x-x_{0})-\Gamma_{M}\dot{x}+\mathcal{P}|\alpha|^{2}
\end{eqnarray}
Here $\Delta=\omega_{L}-\omega_{{\rm cav}}[x=0]$ 
 is the frequency detuning of the incoming laser radiation with respect
 to the optical cavity mode frequency, $\omega'=\partial\omega_{{\rm cav}}/\partial x$ is the derivative with respect to the coordinate ($\omega'=-\omega_{L}/L$  in the usual setup, with $L$ being the cavity length), $\kappa$ is the cavity's intensity ringdown-rate,  $\omega_{M}$ the membrane's mechanical frequency, $x_{0}$  its equilibrium position in the absence of light, and $\Gamma_{M}$ its damping constant.
 The radiation pressure constant $\mathcal{P}$ introduced here has dimensions of frequency squared, and is given by $\mathcal{P}=-(\omega'/\omega_{L})E_{{\rm res}}/m$ , where $m$ is the membrane's effective mass and $E_{{\rm res}}$ is the light energy stored inside the cavity at resonance (proportional
 to the input intensity, $E_{{\rm res}}=4I_{{\rm in}}/\kappa$ for a single-sided cavity).

The steady-state solution  $(\bar{x},\bar{\alpha})$  can be obtained by setting $x(t)=\bar{x}$ and $\alpha(t)=\bar{\alpha}$ and solving the resulting set of nonlinear equations.
 Note that for strong radiation pressure effects, more than one stable solution
 appears (two in the case discussed here).
 This is the static bistability that was found experimentally by Dorsel
 et al. \cite{PhysRevLett.51.1550}
.  We now assume this solution has been found and linearize the equations
 of motion around it, using $|\alpha|^{2}\approx|\bar{\alpha}|^{2}+\bar{\alpha}^{*}\delta\alpha+{\rm c.c.}$

 \begin{eqnarray}
\delta\dot{\alpha} & = & -i\omega'(\bar{x}\delta\alpha+\delta x\bar{\alpha})+i\Delta\delta\alpha-\frac{\kappa}{2}\delta\alpha\\
\delta\ddot{x} & = & -\omega_{M}^{2}\delta x-\Gamma_{M}\dot{x}+\mathcal{P}(\bar{\alpha}^{*}\delta\alpha+{\rm c.c.})+f(t)
\label{eq:dxeq}
\end{eqnarray}
Here we have added a test force leading to an acceleration $f$.
 The response to this force will reveal the change in the membrane's damping
 rate and spring constant brought about by the radiation field.
 At a given driving frequency $\omega$ , we decompose into positive and negative frequency components: $\delta\alpha(t)=\delta\alpha_{-}e^{-i\omega t}+\delta\alpha_{+}e^{+i\omega t}$ , and likewise for $\delta x$  [where $\delta x_{-}=\delta x_{+}^{*}$  due to $\delta x(t)$  being real-valued].
 This leads to
 
 \begin{equation}
\pm i\omega\delta\alpha_{\pm}=-i\omega'(\bar{x}\delta\alpha_{\pm}+\delta x_{\pm}\bar{\alpha})+i\Delta\delta\alpha_{\pm}-\frac{\kappa}{2}\delta\alpha_{\pm},\end{equation}
and therefore $\delta\alpha_{\pm}=\chi_{\alpha}(\pm\omega)\delta x_{\pm}$, with the susceptibility $\chi_{\alpha}$  relating the light response to the membrane motion:

\begin{equation}
\chi_{\alpha}(\omega)=\frac{\bar{\alpha}}{(\Delta-\omega+i\frac{\kappa}{2})/\omega'-\bar{x}}.
\end{equation}
(\ref{eq:dxeq}) leads to

\begin{eqnarray}
-\omega^{2}\delta x_{\pm} & = & -\omega_{M}^{2}\delta x_{\pm}\mp i\omega\Gamma_{M}\delta x_{\pm}+\nonumber \\
 &  & \mathcal{P}(\bar{\alpha}^{*}\delta\alpha_{\pm}+\bar{\alpha}\delta\alpha_{\mp}^{*})+f_{\pm}
 \end{eqnarray}
After inserting  $\delta\alpha_{\pm}$ and using $\delta x_{\pm}=\delta x_{\mp}^{*}$, we find the mechanical response

\begin{equation}
\delta x_{\pm}=\chi(\pm\omega)f_{\pm},
\end{equation}
where the mechanical susceptibility of the membrane has been modified due to the optomechanical coupling:

\begin{equation}
\chi^{-1}(\omega)=\omega_{M}^{2}-\omega^{2}+i\omega\Gamma_{M}+\Sigma(\omega).
\label{eq:OMchi}
\end{equation}
All the novel effects are contained in the optomechanical ``self-energy''

\begin{equation}
\Sigma(\omega)=-\mathcal{P}(\bar{\alpha}^{*}\chi_{\alpha}(\omega)+\bar{\alpha}\chi_{\alpha}^{*}(-\omega))\,.
\end{equation}
The optomechanical damping rate may now be read off from the imaginary part
 of the susceptibility, evaluated at the membrane's resonance frequency:

 \begin{equation}
\Gamma_{{\rm opt}}={\rm Im}[\Sigma(\omega_{M})]/\omega_{M},
\label{eq:Goptgeneral}
\end{equation}
which yields the known result for optical damping:

\begin{equation}
\! \! \! \! \! \! \! \! \! \! \! \! \! \! \! \! \! \! \! \! \! \! \! \! \!  \Gamma_{{\rm opt}}=\omega'\frac{\mathcal{P}}{2\omega_{M}}\left|\bar{\alpha}\right|^{2}\frac{\kappa}{2}\left\{ \frac{1}{[\omega_{M}-\Delta+\bar{x}\omega']^{2}+\left(\frac{\kappa}{2}\right)^{2}}-\frac{1}{[-\omega_{M}-\Delta+\bar{x}\omega']^{2}+\left(\frac{\kappa}{2}\right)^{2}}\right\} 
\end{equation}
which is the difference between the rate of Stokes and anti-Stokes transitions.

The prefactor is equal to $-x_{{\rm ZPF}}^{2}\omega_{R}^{2}\bar{n}\kappa/L^{2}$, where $x_{{\rm ZPF}}^{2}=\hbar/(2m\omega_{M})$.
 Therefore, the optical damping rate is seen to obey the simple formula (see \cite{2007_01_Marquardt_CantileverCooling})

\begin{equation}
\Gamma_{{\rm opt}}=\frac{x_{{\rm ZPF}}^{2}}{\hbar^{2}}[S_{FF}(\omega_{M})-S_{FF}(-\omega_{M})]\,,\end{equation}
where  $S_{FF}$ is the spectrum of radiation pressure force fluctuations.

The damping rate is positive at negative detuning ($\Delta-\omega'\bar{x}<0$), corresponding to cooling, while it is negative at positive detuning,
 leading to an increase in the mechanical quality factor, parametric amplification, and, potentially,
 the onset of self-induced oscillations (once $\Gamma_{{\rm opt}}+\Gamma_{M}<0$) \cite{Marquardt2006}.
 
 Likewise, the shift of the mechanical resonance frequency (optical spring
 effect) is obtained from the real part:
 \begin{equation}
\delta\omega_{M}={\rm Re}[\Sigma(\omega_{M})]/(2\omega_{M})\,.
\label{eq:domGeneral}
\end{equation}

\subsection{Linearized dynamics of dispersively-coupled optomechanical systems}

We now turn to the ``dispersive'' optomechanical device.
 We model it by considering only two modes, residing to the left and to
 the right of the membrane.
 This is the correct description in the limit of a completely reflecting
 membrane.
 We consider the first deviation from that limit, i.e.
 the mode amplitudes $\alpha_{L}$ and $\alpha_{R}$ are coupled by photon tunneling through the membrane, at a frequency  $g$ (the tunneling amplitude).
 When the membrane moves to the right, the frequency of the right mode increases, while that of the left mode decreases.
 They are degenerate at  $x=0$, but the tunneling introduces a splitting and leads to new eigenmodes that
 are symmetric and antisymmetric combinations, as expected for any level
 anticrossing.
 
 These features are incorporated into the following equations of motion:

% Check previous drafts to see the line here that was moved to the top.

\begin{equation}
\dot{\vec{\alpha}}=\M\vec{\alpha}+\left[\begin{array}{c}
\frac{\kappa_{L}}{2}\\
0\end{array}\right],
\label{eq:MIMalphaeq}
\end{equation}
where

\begin{eqnarray}
\vec{\alpha}=\left[\begin{array}{c}
\alpha_{L}\\
\alpha_{R}\end{array}\right];\,\,\,\M & = & \left[\begin{array}{cc}
i(\Delta-\omega'x)-\frac{\kappa_{L}}{2} & -ig\\
-ig & i(\Delta+\omega'x)-\frac{\kappa_{R}}{2}\end{array}\right]
\end{eqnarray}
Again, $\alpha_{L/R}$  have been rescaled such that in the absence of coupling they would reach
 a value of $1$  at resonance for illumination of the left/right cavity (though in the situation
 assumed here, the illumination is only from the left side, as displayed
 by the inhomogeneous term in \ref{eq:MIMalphaeq})). $\Delta$  is the detuning of the laser from the (uncoupled) resonance at $x=0$.
 We have assumed a real-valued tunnel coupling amplitude  $g$.
 Note that the phase of an arbitrary complex amplitude $g$ could be eliminated by incorporating these phases into the definition of $\alpha_{L}$ and $\alpha_{R}$. The optical resonances in the presence of coupling can be found by setting ${\rm det}\M(\Delta)=0$ (with $\kappa_{L}=\kappa_{R}=0$):
 
\begin{equation}
\omega_{{\rm cav},\pm}(x)=\pm\sqrt{g^{2}+(\omega'x)^{2}}
\end{equation}
Comparing this with the general expression for the dispersion, $\omega_{{\rm cav}}(x)=(c/L)\cos^{-1}(r_{d}\cos(4\pi x/\lambda))$, near the degeneracy point and for $r_{d}\rightarrow1$, we find the following relations to the dispersive device's parameters:

 \begin{equation}
g=(c/L)\sqrt{2(1-r_{d})}\,\,\,{\rm and}\,\,\,\omega'=-\omega_{L}/(L/2),
\end{equation}
where it should be noted that $L$ is the full cavity length (comprising both halves), and $\sqrt{2(1-r_{d})}\approx|t_{d}|$ is the transmission amplitude of the membrane.
 Note that in expressions of this kind (like the one for $\omega'$), the optical frequency $\omega_{L}$ is assumed to be that of the original modes at the degeneracy point (small
 deviations do not matter here). The membrane's equation of motion is of the form

\begin{equation}
\ddot{x}=-\omega_{M}^{2}(x-x_{0})-\Gamma_{M}\dot{x}+\mathcal{P}(|\alpha_{L}|^{2}-|\alpha_{R}|^{2})
\end{equation}
For simplicity, we have assumed the two halves of the cavity to be of the
 same length, which is the situation realized in the experiment (otherwise
 one would need to distinguish between $\omega'_{L}$ and  $\omega'_{R}$ as well as $\mathcal{P}_{L}$ and $\mathcal{P}_{R}$). 
 
 Once again, first the steady state solution is found from the system of
 equations
 
\begin{eqnarray}
\bar{\vec{\alpha}} & = & -\frac{\kappa_{L}}{2}\M^{-1}\left[\begin{array}{c}
1\\
0\end{array}\right]\\
0 & = & -\omega_{M}^{2}(\bar{x}-x_{0})-\Gamma_{M}\dot{x}+\mathcal{P}(|\bar{\alpha}_{L}|^{2}-|\bar{\alpha}_{R}|^{2})\,.
\end{eqnarray}
Linearization around this solution and splitting into positive and negative
 frequency components as before leads to

\begin{equation}
\pm i\omega\delta\vec{\alpha}_{\pm}=-i\omega'\sigma_{z}\bar{\vec{\alpha}}\delta x_{\pm}+\bar{\M}\delta\alpha_{\pm},
\end{equation}
where $\bar{\M}$ contains $\bar{x}$ and $\sigma_{z}$ is the Pauli matrix. Thus, we find $\delta\vec{\alpha}_{\pm}=\vec{\chi}_{\alpha}(\pm\omega)\delta x_{\pm}$, with

\begin{equation}
\vec{\chi}_{\alpha}(\omega)=-\omega'[i\omega-\bar{\M}]^{-1}i\sigma_{z}\bar{\vec{\alpha}}\,.
\end{equation}
The mechanical susceptibility, obtained by solving the linearized equation
 for $\delta x$, is analogous to that found for the standard setup, see (\ref{eq:OMchi}), except for containing radiation pressure terms from both the left and
 the right half-cavity:

\begin{eqnarray}
\Sigma(\omega) & = & -\mathcal{P}(\bar{\alpha}_{L}^{*}\chi_{\alpha}^{L}(\omega)+\bar{\alpha}_{L}\chi_{\alpha}^{L*}(-\omega))\nonumber \\
 &  & +\mathcal{P}(\bar{\alpha}_{R}^{*}\chi_{\alpha}^{R}(\omega)+\bar{\alpha}_{R}\chi_{\alpha}^{R*}(-\omega)),
 \label{eq:SigmaMIM}
 \end{eqnarray}
where $\chi_{\alpha}^{L/R}$ refer to the two components of the vector 
$\vec{\chi}_{\alpha}$ The optomechanical damping rate (and the spring effect) can now be obtained
 as before, from (\ref{eq:Goptgeneral}) and (\ref{eq:domGeneral}), by inserting (\ref{eq:SigmaMIM}).
 There is little point in displaying these lengthy expressions explicitly,
 so we will instead discuss the results in terms of plots, for the case
 of  $\kappa_{L}=\kappa_{R}=\kappa$.

\begin{figure}
\includegraphics[width=6.2in]{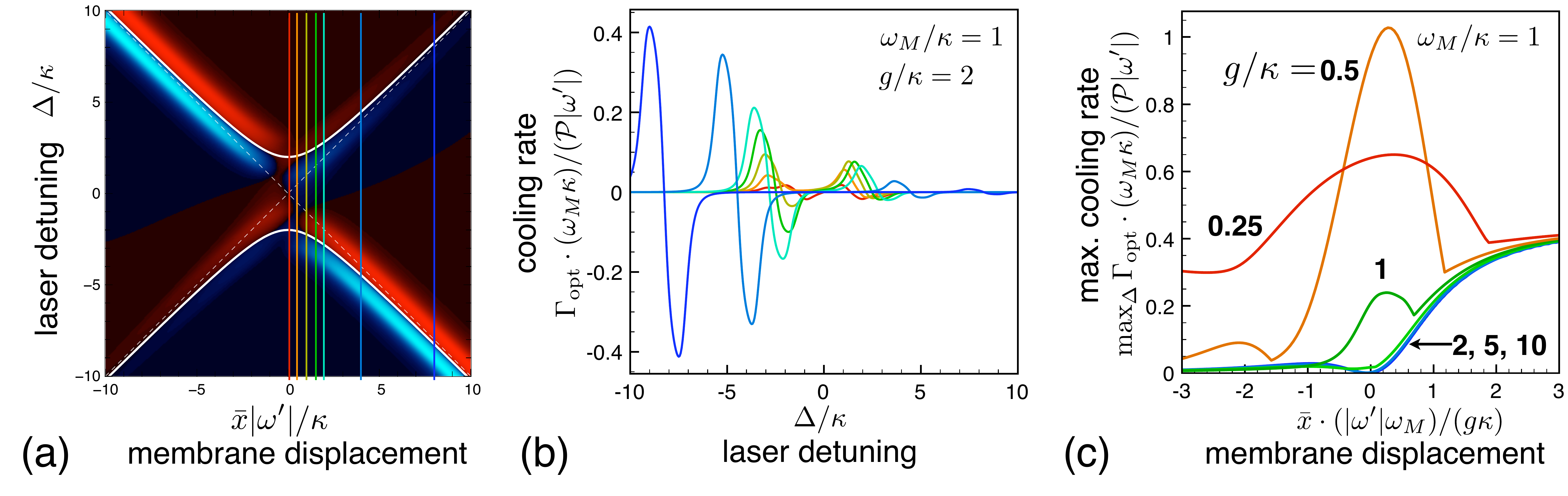}
\caption{(a) The cooling rate $\Gamma_{{\rm opt}}$ for the membrane-in-the-middle setup, as a function of laser frequency
 detuning (with respect to the degeneracy point) and membrane displacement.
 Blue (red) corresponds to cooling (amplification).
 Parameters are $g/\kappa=2$ and $\omega_{M}/\kappa=1$.
 (b) Cross-sections of the preceding plot, taken for several membrane positions: $\bar{x}\cdot|\omega'|/\kappa=0,\,0.5,\,1,\,1.5,\,2,\,4,\,8$.
 (c) Maximizing the cooling rate, as a function of membrane position, for
 various coupling strengths $g$.
 Note that the position has been rescaled by $g$, and the curves coincide for large $g\gg\omega_{M},\,\kappa$.
 They saturate towards large displacements $\bar{x}$, to the value given by the simple theory for the standard setup, while $\Gamma_{{\rm opt}}$
 vanishes at $\bar{x}=0$ (see text) for large $g$.
 For small $g$, the maximum cooling rate can even become larger than the standard limit (see curve labeled  $g/\kappa=0.5$). Note that $\Gamma_{{\rm opt}}$ was maximized over the half-plane $\Delta\leq0$, where the global maximum of $\Gamma_{{\rm opt}}$ is located when $\bar{x}>0$ (the plot would be inverted with respect to $\bar{x}<0$ if we were to maximize over $\Delta\geq0$).
 \label{cross}}
\end{figure}

\subsection{Discussion}

\begin{figure}
\includegraphics[width=7in]{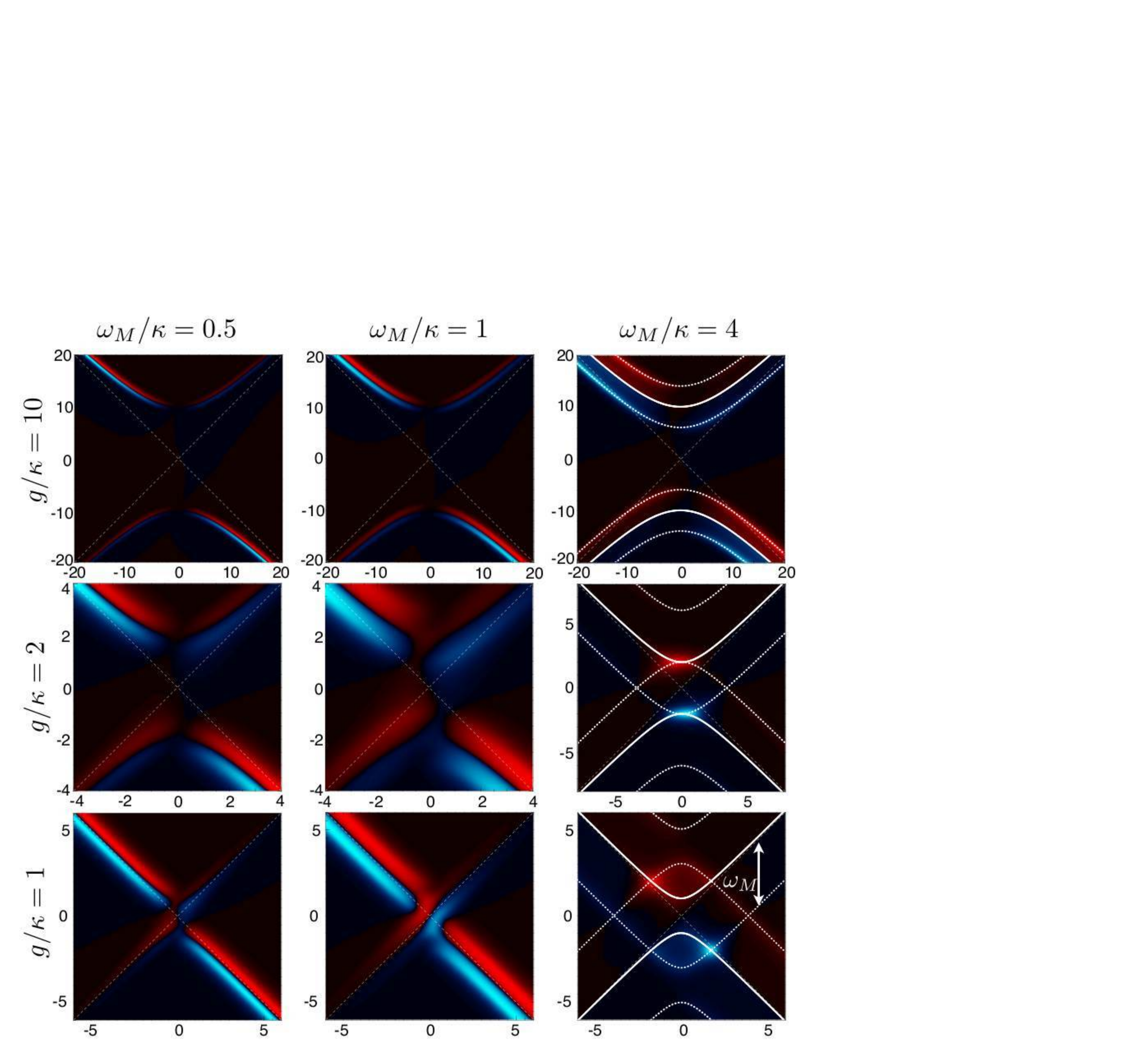}
\caption{The optomechanical cooling rate in the MIM setup near a degeneracy point.
 These plots show $\Gamma_{{\rm opt}}$ as a function of membrane position $|\omega'|\bar{x}/\kappa$ (horiz.) and detuning $\Delta/\kappa$ (vert.), for various values of $\omega_{M}/\kappa$ and $g/\kappa$.
 Blue refers to cooling ($\Gamma_{{\rm opt}}>0$), red to heating or amplification ($\Gamma_{{\rm opt}}<0$).
 \label{figOverview}}
\end{figure}

The diagram of damping rate $\Gamma_{{\rm opt}}$ vs. membrane position $\bar{x}$ and detuning $\Delta$ (both measured in units of the optical resonance width) is shown in figure \ref{cross}. $\Gamma_{{\rm opt}}$ is determined by only two dimensionless parameters. These are the ratio of the membrane frequency $\omega_{M}$ to the cavity ringdown rate $\kappa$ and the ratio of the photon ``tunneling'' rate $g$ to $\kappa$:
 
 \[\frac{\omega_{M}}{\kappa}\,\,{\rm and}\,\,\,\frac{g}{\kappa}\]
 As long as the two optical resonances (i.e.
 the upper and the lower parts of the hyperbolic detuning curve) are separated
 by more than ${\rm max}(\omega_{M},\kappa)$, they can essentially be treated individually.
 In that case, the behaviour of the damping rate $\Gamma_{{\rm opt}}$  in the vicinity of each resonance is qualitatively the same as for a standard ``reflective'' setup.
 That means $\Gamma_{{\rm opt}}$  is positive (negative) for laser light red-detuned (blue-detuned) with
 respect to the resonance, i.e.
 one has cooling (or amplification) for  $\omega_{L}<\omega_{{\rm res}}$  ($\omega_{L}>\omega_{{\rm res}}$).
 When  $\omega_{M}/\kappa$ is small, the maximum $|\Gamma_{{\rm opt}}|$ is reached for a detuning of $\pm\kappa/2$ (the point of maximum slope in the intensity-vs-detuning curve).
 In the resolved-sideband regime $\omega_{M}\gg\kappa$, the maximum is reached at $\pm\omega_{M}$.
 The only quantitative difference is brought about by the change in the slope $\partial\omega_{{\rm res}}/\partial x$, which is directly proportional to the net radiation pressure force
 acting on the membrane.
 As the slope goes to zero near the avoided crossing, so does $\Gamma_{{\rm opt}}\propto(\partial\omega_{{\rm res}}/\partial x)^{2}$. Note that this is a result of our weak coupling approximation and linearization of the equations of motion.  Inclusion of higher order terms would permit two-phonon Raman processes which can lead to cooling for red detunings of $2 \omega_{\rm M}$.
 In addition, the circulating power is smaller when most of the light is
 stored in the right half of the cavity (since we assume illumination from
 the left), and therefore the cooling rate is correspondingly reduced on
 that branch (with positive slope $\partial\omega_{{\rm res}}/\partial x$), as can be seen in figure \ref{figOverview}.
 On the amplification side ($\Gamma_{{\rm opt}}<0$), the membrane may settle into a state of self-sustained oscillations
 when $\Gamma_{{\rm opt}}+\Gamma_{M}$ becomes negative.
 Those regions of instability can therefore directly be read off diagrams
 such as those in figure \ref{figOverview}, once the mechanical damping rate $\Gamma_{M}$ is given.  

When the resonances touch, i.e. when they get closer than ${\rm max} (\omega_{M},\kappa)$,
the regions of cooling and amplification become visibly distorted,
with intricate patterns as a result. In any case however, the diagrams
remain inversion symmetric around the degeneracy point (upon  
simultaneous
change of the sign of $\Gamma_{{\rm opt}}$).

The physics of this regime can best be understood by analyzing the
cases where the mechanical frequency $\omega_{M}$ becomes comparable
to or even larger than the splitting $2g$ of the dispersion relation,
while the ring-down rate remains small. For the following discussion,
we therefore refer the reader to the lower right panel of figure  
\ref{figOverview}
($g/\kappa=1$ and $\omega_{M}/\kappa=4$). As seen in that figure,
the cooling or heating rate is apparently maximal at places where
the incoming radiation is in resonance either with the optical  
eigenfrequencies
$\omega_{{\rm cav},\pm}(x)$, or with their sidebands $\omega_{{\rm  
cav},\pm}(x)\pm\omega_{M}$
(full and dashed lines in that panel). The rate becomes particularly
pronounced when these dispersion curves cross. At these places, there
is interference between the eigenmode that is nearby in frequency,
and the excitation of the other eigenmode via Raman scattering. Indeed,
this interference is necessary to explain the remarkable fact that
there can be some cooling or heating even at $x=0$, particularly
when $2g$ becomes smaller than $\omega_{M}$: At the degeneracy point
$x=0$, we have the eigenmodes as symmetric and antisymmetric  
combinations
of the basis modes, $\alpha_{\pm}=(\alpha_{L}\pm\alpha_{R})/\sqrt{2}$.
Consequently, the radation pressure force

\begin{equation}
F_{{\rm rad}}\propto\left|\alpha_{L}\right|^{2}-\left|\alpha_{R}\right| ^{2}\propto\left|\alpha_{+}+\alpha_{-}\right|^{2}-\left|\alpha_{+}- \alpha_{-}\right|^{2}\end{equation}
vanishes identically, unless there is interference between $\alpha_{+}$
and $\alpha_{-}$. This can happen at the points where the dispersions
$\omega_{\pm,{\rm cav}}(x)$ and $\omega_{\pm,{\rm cav}}(x)\pm\omega_{M}$
cross. It also happens elsewhere, to a lesser extent, due to the
finite cavity ring-down rates $\kappa_{L,R}$, which broaden the  
resonances.
The strongest effect is observed when $\omega_{M}=2g$, where the
resonance conditions are fulfilled simultaneously right at the  
degeneracy
point (see Fig.~\ref{figOverview}, panel with $g/\kappa=2,\omega_{M}/ \kappa=4$;
and Fig.~\ref{cross}(c), $g/\kappa=0.5,\,\omega_{M}/\kappa=1$).
On the other hand, the effect vanishes in the limit $\omega_{M}/g \rightarrow0$.
In that limit, the standard picture is sufficient, when taking into
account the suppression of the slope $\partial\omega_{{\rm cav}}(x)/ \partial x$
at the degeneracy point, which reduces the cooling and heating rates
to zero.

In summary, in a certain regime it is possible to cool the membrane
even at $x=0$ (i.e., where the bands are flat). This discussion may
be important for experiments in which the membrane is kept at $x=0$
to realize a phonon QND measurement, as described in Ref.  
\cite{Harris08}.

% Aash's contribution.
\section{Signatures of quantum behaviour in a weak energy measurement}

As has been discussed above and elsewhere\cite{Harris08}, a dispersive optomechanical device can be operated in a regime where the cavity frequency depends directly on $x^2$, the position-squared of a macroscopic mechanical oscillator.  As a result, one can make a direct measurement of the oscillator's energy $E = \hbar \omega_M ( n + 1/2)$ where $n$ is the number of phonons in the membrane \cite{Harris08}.
One drives the cavity on resonance and measures the phase of the transmitted beam; this phase is proportional to $E$.  We thus have the possibility to detect a truly quantum aspect of the oscillator: the quantization of its energy.  Note that this is impossible to do with a linear position detector (e.g.~a cavity whose frequency depends directly on $x$), as in this case one measures both the energy {\it and} phase of the oscillator, and is thus subject to the usual limitations imposed by quantum back-action \cite{Braginsky80}.  All recent experiments in optomechanics and electromechanical systems have (to the best of our knowledge) employed linear position detectors, and are hence subject to these limitations.

While having a non-linear coupling to the oscillator is certainly a prerequisite to detecting the quantum nature of its energy, it is not in itself enough:  one also needs to consider the output noise of the detector.  Here, this output noise consists of the shot noise in the transmitted beam through the cavity, plus any additional technical noise associated with determining the phase of this beam.  If this measurement was truly QND, this output noise would not be a problem: one could achieve any desired sensitivity by simply averaging the output signal for a sufficiently long time \cite{Braginsky80}.  The
back-action of a perfect QND measurement does not affect the measured observable, and thus the oscillator's energy would not fluctuate during the measurement.  However, this is not the case for any real experiment, which is performed at some small but non-zero temperature.  Because of this small temperature and the non-zero oscillator damping, the oscillator's energy will indeed fluctuate if one waits long enough.  There is thus a limit to how long one can average, and thus to how well one can resolve the quantum nature of the oscillator's energy.  

In this section, we will address quantitatively the limitations on detecting quantum behaviour 
arising from the combination of the weak nature of the measurement and the unavoidable (thermal) fluctuations in the oscillator energy.  We will focus on the realistic case where one can only obtain an energy resolution corresponding to a single quanta after averaging for a time comparable to (or longer than) the lifetime of a phonon Fock state.  As such, one is no longer truly measuring the instantaneous energy of the oscillator; instead, one is measuring the time-integrated energy fluctuations of the oscillator.  We will calculate this quantity for both a classical and a quantum dissipative oscillator, and will discuss whether differences between the two are experimentally resolvable.  We will focus throughout on the experimental conditions proposed in Ref.\cite{Harris08}.  In particular, we assume a situation where the oscillator is initially near its ground state, but is coupled to a dissipative bath with  a temperature $T_{\rm bath} \gg \hbar \omega_M / k_B$.  In the proposed experiment this is realized by laser-cooling the membrane to its ground state and then shutting off the cooling beam while an energy measurement is made.  Note that the somewhat related situation of QND measurement of qubit energy was studied theoretically in Ref.~\cite{Gambetta07}.

\subsection{Model and measurement sensitivity in the zero-damping limit}

The quantity measured in the experiment is the phase shift of the transmitted beam through the cavity (or, equivalently, the error signal in a Pound-Drever-Hall setup); by dividing out a proportionality factor (the ``gain" of the measurement), one can refer this signal back to the mechanical oscillator, expressing it as a number of quanta $\tn(t)$:
\begin{eqnarray}
	\tn(t) = n(t) + \xi(t)
\end{eqnarray}
where $n(t)$ is the actual number of oscillator quanta, and $\xi(t)$ is the added noise of the measurement.  We take $\xi(t)$ to be Gaussian white noise with a (two-sided) spectral density $S_{nn}$, i.e.:
\begin{eqnarray}
	\langle \xi(t_1) \xi(t_2) \rangle = S_{nn} \delta(t_1 - t_2)
\end{eqnarray}
For the cavity system and a shot-noise limited Pound-Drever-Hall measurement, one has:
\begin{eqnarray}
	S_{nn} = \frac{\hbar c \lambda^3 (1-r_c) }{4096 \pi F^2 P_{\rm in} x_{\rm m}^4} 
\end{eqnarray}
Note that $S_{nn}$ depends both on the amount of output noise in the measurement, and on the strength of the cavity - oscillator coupling.

We will be interested throughout in the case of a weak measurement, where one must time-average the output signal to counteract the effects of the added noise.  We are thus led to the quantity $\tm(t)$, the time-integral of the inferred number of quanta $\tn(t)$:
\begin{eqnarray}
	\tilde{m}(\tavg) =   \int_0^{\tavg} dt' \tn(t')
		& = & \int_0^{\tavg} dt' n(t') + \int_0^{\tavg} \xi(t') \nonumber \\
	 & \equiv & m(\tavg) + \int_0^{\tavg} \xi(t')
\end{eqnarray}
As the added noise is white, the probability distribution $\Pmeas(\tm,\tavg)$ for the measured quantity $\tm$ is a simple convolution of a Gaussian and the probability distribution $P(m)$ of the true oscillator energy fluctuations $m$:
\begin{eqnarray}
	% better way to do the below?
	\! \! \! \! \! \! \! \! \! \! \! \! \! \! \! \! \! \! \Pmeas(\tm,\tavg) = \int dm' \left[ P(\tm-m',\tavg)\cdot
	\frac{1}{\sqrt{2 \pi \left( \sigma_n \tavg \right)^2}}
		\exp \left( - 
			\frac{ (m'/\tavg  )^2 }
				{ 2 \sigma_n ^2 } 
			\right) \right]
	\label{eq:Pmeas}
\end{eqnarray}
with
\begin{eqnarray}
	\sigma_n(\tavg) = \sqrt{ \frac{S_{nn}}{\tavg} }
\end{eqnarray}

Consider first the ideal case where the measurement is completely back-action free, and where the oscillator damping $\gamma \ra 0$, meaning that there are no thermal energy fluctuations.  
In this case, if the oscillator starts with $n_0$ quanta, it will always have $n_0$ quanta:  $P(m,\tavg) = \delta(m - n_0 \tavg)$.  The distribution of $\tm$ is then just a Gaussian:
\begin{eqnarray}
	\Pmeas(\tm,\tavg) = \frac{1}{\sqrt{2 \pi \left( \sigma_n \tavg \right)^2}}
		\exp \left( - 
			\frac{ (\tm/\tavg - n_0  )^2 }
				{ 2 \sigma_n ^2 } 
			\right)
\end{eqnarray} 
To see evidence of the oscillator's quantum nature, we would like to be able to resolve Fock states that differ by a single quanta.  For $\gamma=0$, these two states will each give Gaussian distributions of $\tilde{m}$ having means separated by $\tavg$.  As is standard, we can describe the distinguishability of these two Gaussians by a signal to noise ratio $R_{SNR}(\tavg)$.  This is simply the ratio of the signal power to the noise power:
\begin{eqnarray}
	R_{SNR}(\tavg) = 
		\frac{ \left[ \langle \tm \rangle_1 - \langle \tm \rangle_2 \right]^2 }	
			{\left( \Delta \tm_1 + \Delta \tm_2 \right)^2 } = 
	\frac{\tavg^2}{ \left(2 \tavg \sigma_n \right)^2 } = 
		\frac{\tavg}{4 S_{nn} }
\end{eqnarray}
where $\Delta \tm_1$ denotes the standard deviation of $\tm$ for the first Gaussian distribution, etc.  As expected, $R_{SNR}(\tavg)$ can be made arbitrarily large by increasing the averaging time $\tavg$.  In particular, the two Gaussians become resolvable (i.e.~the averaged distribution has two as opposed to one maximum) when $R_{SNR}(\tavg) \geq 1$\footnote{Note that our definition of the SNR ratio is smaller by a factor of two than the SNR ratio $\Sigma$ used in Ref.~\cite{Harris08}} .  

\subsection{Distinguishing quantum from classical when $\gamma >0$}

The story becomes somewhat more complicated when we now include the unavoidable fluctuations of $n(t)$.  We will consider the experimentally relevant case where these fluctuations are {\it only} due to the dissipative bath coupled to the oscillator, and not to the back-action of the measurement.  As discussed in Ref.~\cite{Harris08}, there is a small back-action effect associated with the fact that the cavity is coupled to $x^2$ and not the oscillator energy; this however is a much weaker effect than the thermal fluctuations we consider.    The thermal bath coupled to the oscillator will  cause a given oscillator Fock state $| n \rangle$ to decay at a rate $\Gamma_n$.  A simple golden rule calculation yields:
\begin{eqnarray}
	\Gamma_n = \gamma \left[ \nbar + n (2 \nbar + 1) \right]
\end{eqnarray}
where $\gamma$ is the damping rate of the oscillator,
\begin{eqnarray}
	\nbar = \left( \exp\left[ \frac{\hbar \omega_M}{k_B \Tbath} \right] -1 \right)^{-1},
	\label{eq:nbar}
\end{eqnarray}
 and $\Tbath$ is the bath temperature.
 
 Due to these thermal fluctuations, the distribution of the measured quantity $\tm$ will not be Gaussian.  To obtain a very rough estimate of whether our measurement can still resolve quantum energy behaviour, we could still attempt to use the SNR ratio derived above; this was the approach taken in \cite{Harris08}.  We assume that we start in the ground state (to maximize the Fock state lifetime $1/\Gamma_n$), and use an averaging time equal to the lifetime of this state.   Letting $\tau = 1 / \Gamma_0$ represent the lifetime of the ground state, we thus have as an approximate figure of merit
:
\begin{eqnarray}
	R \equiv R_{SNR}(\tavg = \tau) = \frac{ \tau }{4 S_{nn}} 
		= 
		\frac{ 1}{4 \gamma \nbar S_{nn}}
		\label{eq:R} 
\end{eqnarray}
One might guess that if $R > 1$, one can resolve quantum aspects of the oscillator's energy;  in \cite{Harris08}, it was shown that achieving $R \sim 1$ could be possible in the next generation of experiments.   However, the condition $R > 1$ is clearly an approximate one, as it neglects all the complexities arising from the fluctuations of the oscillator.  In particular, the two distributions one is trying to distinguish are not Gaussian, and thus it is by no means clear that the SNR ratio $R$ will remain a good measure of distinguishability.

We will now assess more accurately the conditions required to resolve quantum-classical differences.  The first step will be to ignore the added noise of the detector, and focus on the probability distribution of the ``true" time-integrated oscillator energy $m$.  We will do this in both the cases of a classical oscillator and a quantum oscillator; the respective distributions will be denoted $P_{\rm cl}(m,\tavg)$ and $P_{\rm q}(m,\tavg)$.  Having these distributions, we will then add the effects of the added noise $S_{nn}$, and ask whether the corresponding {\it measured} distributions $\Pmeascl(m,\tavg)$ and $\Pmeasq(m,\tavg)$ (as given by (\ref{eq:Pmeas})) are distinguishable for a given level of noise and averaging time.  As the distributions involved will be non-Gaussian, we will need to use a more sophisticated measure of distinguishability than the signal-to-noise ratio $R$ used in the Gaussian case.  We will make use of an information-theoretic measure, the accessible information $\II$.

We start with the first step of our program:  what are the probability distributions of the true integrated oscillator energy fluctuations $m$?  Given the relative weakness of cavity-oscillator couplings, we will necessarily need to use averaging times comparable to or even longer than the lifetimes of oscillator Fock states.  As a result, the experiment is no longer about measuring the instantaneous energy of the oscillator.  Rather, we are asking whether one can see quantum behaviour in the energy fluctuations of the oscillator.  The quantities we we wish to calculate ($P_{\rm q}(m,\tavg)$ and $P_{\rm cl}(m,\tavg)$) are thus formally analogous to the well-studied full counting statistics of charge in mesoscopic electron systems \cite{Levitov03}; there, one wishes to calculate the statistics of the time-integrated current through a mesoscopic conductor.  Given this similarity, we can employ a similar calculational technique in our problem.  This was essentially done in Ref.~\cite{Clerk07a}, where the motivation was to describe an experiment where a qubit is used to detect Fock states in a nanoresonator.  One calculates the dephasing of a qubit whose energy is directly proportional to the energy of a dissipative oscillator; this immediately yields the generating function of $P_{\rm q}(m,\tavg)$, $\tP_{\rm q}(\lambda,\tavg)$, defined by:
\begin{eqnarray}
	\tP_{\rm q}(\lambda,\tavg) = \int_{-\infty}^{\infty} dm e^{-i \lambda m} P_{\rm q}(m,\tavg)
\end{eqnarray}
In Ref. \cite{Clerk07a}, the focus was to understand the time-dependence of $\tP_{\rm q}(\lambda,\tavg)$, and hence the dephasing spectrum of the qubit.  Here, the focus will instead be on its $\lambda$ dependence, as the Fourier transform of $\tP_{\rm q}(\lambda,\tavg)$ will yield the desired distribution $P_{\rm q}(m,\tavg)$.  

Consider the initial condition corresponding to the proposed experiment:  the oscillator is initially in a thermal state corresponding to a temperature $\Tinit$ which differs from the bath temperature $\Tbath$.  Using the method of Ref. \cite{Clerk07a}, one finds that the corresponding generating function for a quantum oscillator is given by:
\begin{eqnarray}
	\tP_{\rm q} (\lambda,t) & = &
		e^{ \gamma t / 2} e^{-i  (\alpha-\lambda) t / 2}
		\frac{1 - M}{1 - M e^{-i \alpha t} }
	\label{eq:Pq}
\end{eqnarray}
with
\numparts
\begin{eqnarray}
	\alpha & = & 
		\sqrt{ \left(\lambda - i \gamma \right)^2 
		- 4 i \lambda \gamma \nbar  } 
	\label{eq:AlphaEqn} \\
	M & = &
		\frac{
			2 \lambda \ninit  - 
				\left( \alpha -  \lambda +   i \gamma \right)
			}
			{
			2 \lambda \ninit  + 
				\left( \alpha + \lambda -  i \gamma \right)
			}		
		\label{eq:MEqn}
\end{eqnarray}
\endnumparts
%Note that at $T=0$, we have simply $\alpha = \lambda - i \gamma / \omega_M$ and $M=0$.
Here, $\nbar$ is the Bose-Einstein factor associated with the bath temperature 
(cf. (\ref{eq:nbar})), while $\ninit$ is the Bose-Einstein factor associated with the inital oscillator temperature $T_{\rm init}$.  
%Using the above, we can calculate the distribution of the experimental output, $\Pmeas(\tm)$, for various values of temperature.

For comparison purposes, we also require $P_{\rm cl}(m,\tavg)$,  the distribution of $m$ for a classical dissipative oscillator.  To obtain this, we use $\tP_{\rm q}(\lambda,\tavg)$ in (\ref{eq:Pq}) to find $P_{E,{\rm q}}(s,\tavg)$ the distribution of integrated oscillator {\em energy} fluctuations $s = \hbar \omega (m + \tavg/2)$; this involves a simple change of variables.  It is then straightforward to take the classical $\hbar \ra 0$ limit to find $P_{E,{\rm cl}}(s,\tavg)$.  Defining the corresponding generating function via:
\begin{eqnarray}
	\tP_{E,{\rm cl} }(\chi,\tavg) = \int ds e^{-i \chi s} P_{E,{\rm cl} }(s,\tavg)
	\label{eq:PEcl}
\end{eqnarray}
we find:
\begin{eqnarray}
	\tP_{E,{\rm cl} }(\chi,t) & = &
		e^{ \gamma t / 2} e^{-i  \alpha_{cl} t / 2}
		\frac{1 - M_{\rm cl} }{1 - M_{\rm cl} e^{-i \alpha t} }
\end{eqnarray}
with
\begin{eqnarray}
	\alpha_{\rm cl} & = & 
		\sqrt{ - \gamma ^2 
		- 4 i k_B \Tbath \chi \gamma   } 
	\label{eq:AlphaEqn2} \\
	M_{\rm cl} & = &
		\frac{
			2 k_B \Tinit \chi  - 
				\left( \alpha_{\rm cl}  +   i \gamma \right)
			}
			{
			2 k_B \Tinit \chi   + 
				\left( \alpha_{\rm cl}  -  i \gamma \right)
			}		
%			\\
%	M & = &
%		\frac{
%			2 \lambda e^{-\beta \omega_M} + \left( e^{-\beta \omega_M}-1 \right)
%				\left( \alpha - \left( \lambda -  \frac{ i \gamma}{ \omega_M} \right) \right)
%			}
%			{
%			2 \lambda + \left( e^{-\beta \omega_M}-1 \right)
%				\left( \lambda - \alpha + \frac{ i \gamma}{ \omega_M} \right)
%			}		
		\label{eq:MEqn}
\end{eqnarray}
The corresponding classical distribution of $m$, $P_{\rm cl}(m,t)$, follows from:
\begin{eqnarray}
	\tP_{\rm cl}(\lambda,t) = P_{E,{\rm cl} }\left(
		\chi = \frac{\lambda }{ \hbar \omega_M } ,t \right)
		e^{i \lambda t / 2}
	\label{eq:Pclass}
\end{eqnarray}

While the form of the classical generating function $\tP_{E,cl}(\chi,t)$ may seem unfamiliar, it is easy to check its behaviour in some simple limits.  For example, consider the limit where $\gamma \ra 0$.  Equation (\ref{eq:PEcl}) then yields:
\begin{eqnarray}
	\tP_{E,{\rm cl}}(\chi) \ra \frac{1}{i \chi k_B \Tinit + 1}
\end{eqnarray} 
which corresponds to a simple Boltzman distribution as expected:
\begin{eqnarray}
	P_{E,{\rm cl}}(s,t) = \frac{1}{k_B \Tinit t} \exp \left(
		-\frac{(s/t)}{k_B \Tinit } \right)
\end{eqnarray}

We now have analytic expressions for the distribution of time-integrated energy fluctuations of both a classical oscillator ($P_{\rm cl}(m,t)$, cf. (\ref{eq:Pclass})) and a quantum oscillator ($P_{\rm q}(m,t)$, cf. (\ref{eq:Pq})).  The corresponding distributions of the measured quantity $\tilde{m}$ can be easily found by including the effects of the added noise $S_{nn}$ via (\ref{eq:Pmeas});
we denote these (respectively) as $\Pmeascl(m)$ and $\Pmeasq(m)$.  To assess how different these two (non-Gaussian) distributions are from one another, we will consider their mutual information $\II$.  This measure of distinguishability is defined as \cite{CoverandThomas}:
\begin{eqnarray}
	\II[P_1,P_2] = H[ (P_1 + P_2) / 2 ] - 
		\frac{1}{2} \left(
			H[ P_1 ] + H[P_2] 
		\right)
\end{eqnarray}
where $H[P]$ is the Shannon entropy of the distribution $P$:
\begin{eqnarray}
	H[P] := \int dm P(m) \log_2 P(m)
\end{eqnarray}
We can interpret the first term in Eq. (50) as the information in a signal in which each instance is drawn randomly from either $P_1$ or $P_2$.  
The second term is the average information under the same circumstances except that we are told from which distribution the signal is drawn.  If telling us which distribution was used makes no difference then the two distributions are identical and the mutual information is zero. The larger the value of the mutual information $\II[P_1, P_2]$, the more distinguishable are the two distributions $P_1$ and $P_2$.  $\II$ is a convenient measure both because it is applicable to arbitrary distributions $P_1$ and $P_2$, and because it has a direct information-theoretic interpretation in terms of information transmission rates down noisy communication channels \cite{CoverandThomas}.  We note for two Gaussians distributions with identical standard deviations, $\II \simeq 0.49$ when $R_{SNR}=1$, while $\II \ra 1$ for $R_{SNR} \gg 1$.
 
\subsection{Results}

Equations (\ref{eq:Pmeas}), (\ref{eq:Pq}) and (\ref{eq:Pclass}) can now be used to quantitatively 
assess whether quantum versus classical differences can be resolved under a variety of different experimental conditions. In what follows, we will present only a few selected results relevant to the experiment proposed in Ref.~\cite{Harris08}.

\subsubsection{Measurement runs starting in the oscillator ground state}

\begin{figure}
\begin{center}
\includegraphics[width=6.0in]{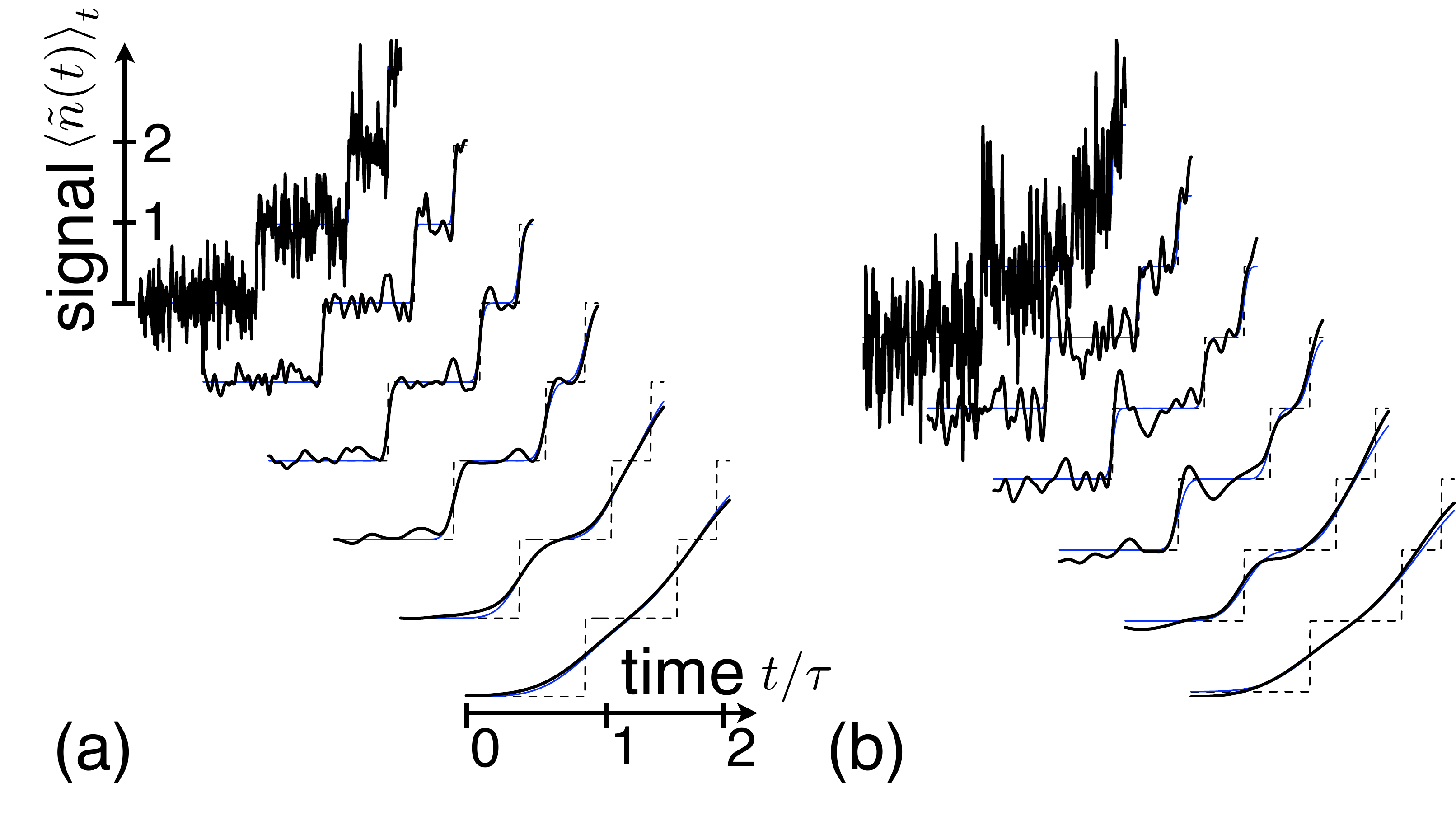}
\caption{Quantum jump traces in the presence of noise and temporal averaging, for different noise     
strengths ((a) and (b)) and increasing averaging time (top to bottom).                       
The plots show the traces $\left\langle{\tilde n}(t)\right\rangle_{t}$ that would            
be observed by doing a sliding time-average of $n(t)$, including the noise $\xi(t)$.         
Here the time-average was done by convoluting with a Gaussian whose width is set             
by ${\tilde t}_{\rm avg}$, in such a way as to have the simple relation                      
$\left\langle\left\langle\xi\right\rangle_t^2\right\rangle=S_{nn}/{\tilde t}_{\rm avg}$.     
The noise strength has been chosen as $S_{nn}/\tau=0.001$ and $S_{nn}/\tau=0.004$ in (a) and      
(b), respectively.                                                                           
The time-interval displayed in each curve is $2\tau$, where $\tau$ is the                    
ground state lifetime. From top to bottom, successive curves arise from the                  
same trace by averaging over increasing time-intervals:                                      
${\tilde t}_{\rm avg}/\tau=0.01, 0.05, 0.1, 0.2, 0.5, 1.0$. Curves have been                 
displaced horizontally and vertically for clarity. The values of                             
a signal-to-noise ratio, defined in correspondence to the discussion in the main text,       
are (from top to bottom): (a) ${\tilde t}_{\rm avg}/(4 S_{nn}) =$                            
$2.5, 12.5, 25, 50, 125, 250$                                                                
and (b) $0.625, 3.125, 6.25, 12.5, 31.25, 62.5$.
\label{trajectories}}
\end{center}
\end{figure}

\begin{figure}
\begin{center}
\includegraphics[width=6.0 in]{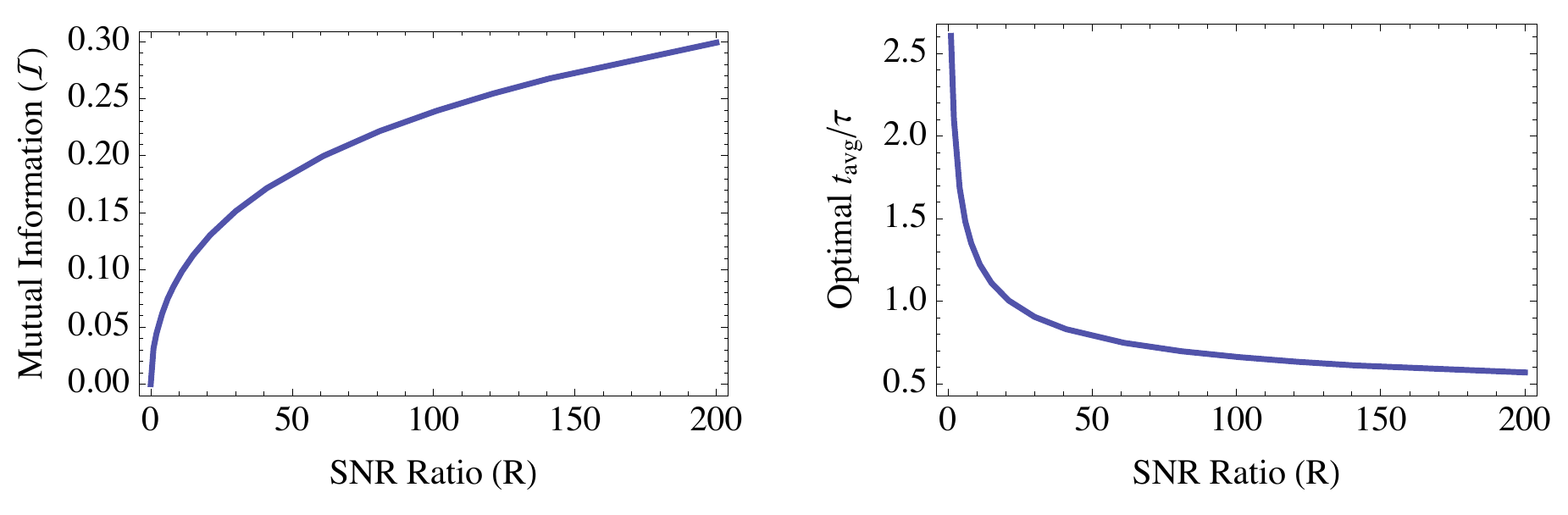}
\caption{Left: distinguishability of quantum versus classical energy fluctuations, as measured by the mutual information $\II[ P_{\rm meas,q}, P_{\rm meas,cl}]$, versus the inverse added noise $R = 1 / (4 S_{nn} \tau)$.  In each case, we have used an optimal averaging time $\tavg$ which maximizes $\II$ and have started the oscillator in the ground state $\ninit = 0$.  Following Ref.~\cite{Harris08}, we have taken 
$\omega_M / 2 \pi = 10^5 {\rm Hz}$, $\gamma/ \omega_M = 1.2 \times 10^{-7}$ and $\Tbath = 300 { \rm mK}$.  Finally, we have shifted the quantum distribution so that both the quantum and classical distributions have the same mean, as in experiment, this shift of the mean would be hard to detect.  Note that an SNR ratio of $R=1$ in the Gaussian case corresponds to $\II \simeq  0.49$.  Right:  optimal averaging time versus $R$, same parameters.} 
\label{fig:GndStateIIPlot}
\end{center}
\end{figure}

\begin{figure}
\begin{center}
\includegraphics[width=6.5 in]{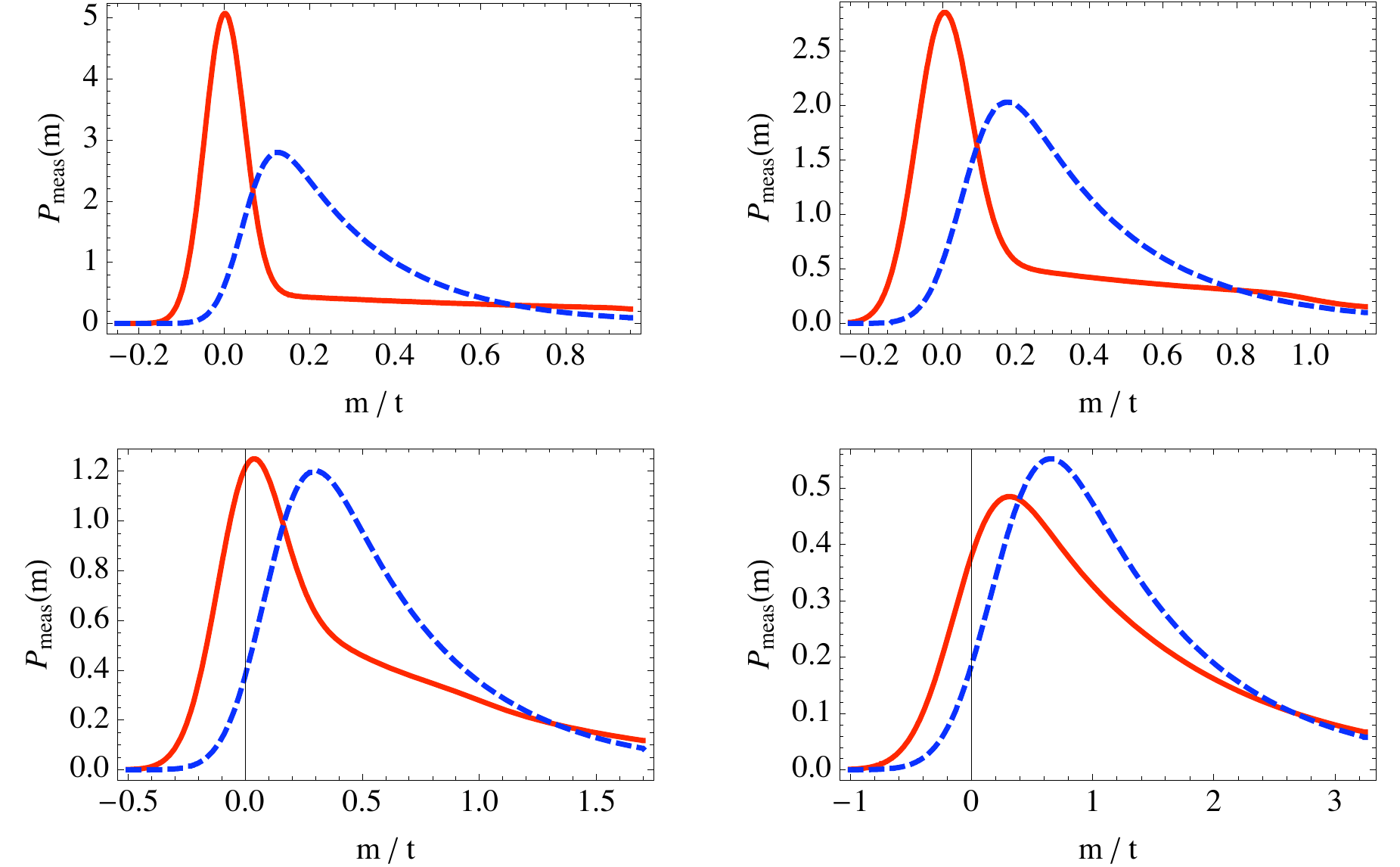}
\caption{Distributions of the integrated output of the experiment, 
for both the cases of a classical oscillator 
($P_{\rm meas,cl}(m)$, dashed blue) and a quantum oscillator 
($P_{\rm meas,q}(m)$, solid red) oscillator.  Following Ref.~\cite{Harris08}, we have taken $\omega_M / 2 \pi = 10^5 {\rm Hz}$, $\gamma/ \omega_M = 1.2 \times 10^{-7}$ and $\Tbath = 300 {\rm mK}$.
In each case, the oscillator starts in the ground state, and an optimal averaging time has been used; we have also shifted the quantum distribution in each case to remove the zero-point shift in the average.  The four panels correspond to different levels of added noise $S_{nn}$:  $R=200$ (top left), $R=61$ (top right), $R=11$ (bottom left) and $R=1$ (bottom right).  In each plot, the range of $m$ has been chosen to display $90\%$ of the 
area of the quantum curve. }
\label{fig:GndStateDist}
\end{center}
\end{figure}

We first consider the ideal situation where the oscillator has been cooled to its ground state: $\Tinit = \ninit = 0$.  The cooling beam is then shut off, and the number-state measurement is made.  During this time, the oscillator rapidly heats up due to its coupling to the equilibrium heat bath at temperature $\Tbath$.  We will focus on the experimentally relevant case where $\gamma \ll \omega_M$, $\nbar \gg 1$, and on averaging times small enough that the average number of quanta in the oscillator remains much smaller than $\nbar$.  In this regime, the oscillator damping $\gamma$ and the bath temperature $\Tbath$ essentially only enter via the time-scale $\tau$, the lifetime of the $n=0$ Fock state.  It is this time-scale which determines the initial heating-up of the oscillator:
\begin{eqnarray}
	\langle \hat{n}(t) \rangle & = & 
		 \nbar (1-e^{-\gamma t}) \\
		& \simeq &
		\gamma \nbar t = t / \tau
\end{eqnarray}
Given these conditions, there are two relevant questions.  First, given a certain noise level $S_{nn}$, what is the optimal averaging time?  Second, given that we have optimized the averaging time, how does the distinguishability depend on $S_{nn}$?

To illustrate this process, we show a series of calculated time traces corresponding to such a scenario in figure \ref{trajectories} for various values of the averaging time. In the left panel of figure \ref{fig:GndStateIIPlot}, we show how the measurable distinguishability between the classical and quantum distributions depends on the noise level $S_{nn}$, as parameterized by $R$ (cf.(\ref{eq:R})).  The distinguishability is measured by the mutual information $\II$ between the expected experimental distributions for a classical and quantum oscillator ($\Pmeascl(m)$ and $\Pmeasq(m)$ respectively).  Each run corresponds to starting the oscillator in the ground state and using an averaging time which maximizes $\II$; the value of the averaging time is shown in the right panel of figure \ref{fig:GndStateIIPlot}.  Note that before computing the mutual information $\II$, we have shifted the quantum distribution to remove the zero-point energy contribution (as resolving this difference in an experiment would be very difficult); the result is that both the classical and quantum distributions have identical means.  We see that in general, one needs a noise level small enough that $R \gg 1$ to unambiguously resolve classical-quantum differences: the simple Gaussian estimate which suggests $R \sim 1$ is sufficient is too optimistic.  Classical and quantum distributions of $m$ for different values of $R$ are shown in figure \ref{fig:GndStateDist}.   

\subsubsection{Starting at a finite temperature}

It is also interesting to ask what happens if the oscillator does not start in the ground state (i.e. $n_{\rm init} > 0$).  In practice, one might not be able to cool the oscillator all the way to the ground state.  
Even if one could cool to the ground state, it would be very useful experimentally to extract as much information as possible in one run of the experiment.  We saw above that if the oscillator starts in the ground state, the optimal averaging time is on the order of $\tau$.  After this initial averaging time, the average number of quanta in the oscillator will be $\sim 1$.  One could imagine starting a second averaging period at this point; the question is whether the initial temperature of the oscillator will make quantum versus classical differences even harder to see.

\begin{figure}
\begin{center}
\includegraphics[width=6.5 in]{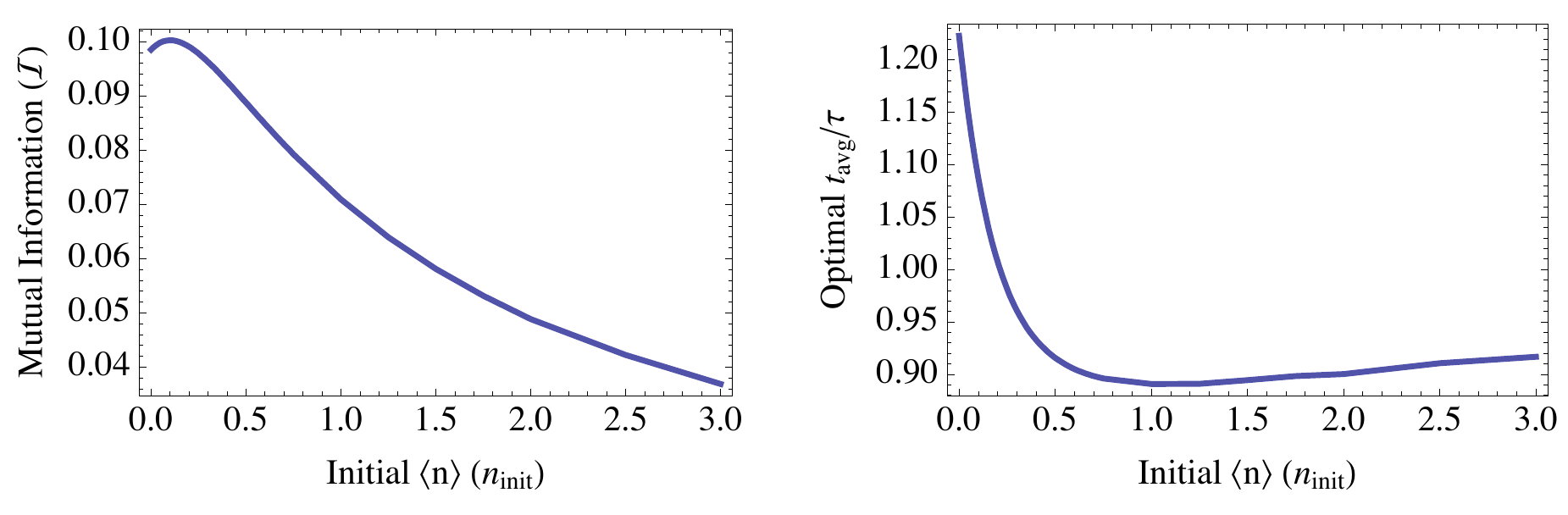}
\caption{Left: distinguishability of quantum versus classical energy fluctuations (as measured by $\II[ P_{\rm meas,q}, P_{\rm meas,cl}]$), versus initial oscillator temperature, and for a fixed noise strength corresponding to $R=11$.
Following \cite{Harris08}, we have taken $\omega_M / 2 \pi = 10^5 {\rm Hz}$, $\gamma/ \omega_M = 1.2 \times 10^{-7}$ and $\Tbath = 300 {\rm mK}$.  In each case we have used an optimal averaging time, and have shifted each quantum distribution so that both the quantum and classical distributions have the same mean.  Right: optimal averaging time as a function of $n_{\rm initial}$ for the same choice of parameters. } 
\label{fig:ThermalSNR11IIPlot}
\end{center}
\end{figure}

\begin{figure}
\begin{center}
\includegraphics[width=6.5 in]{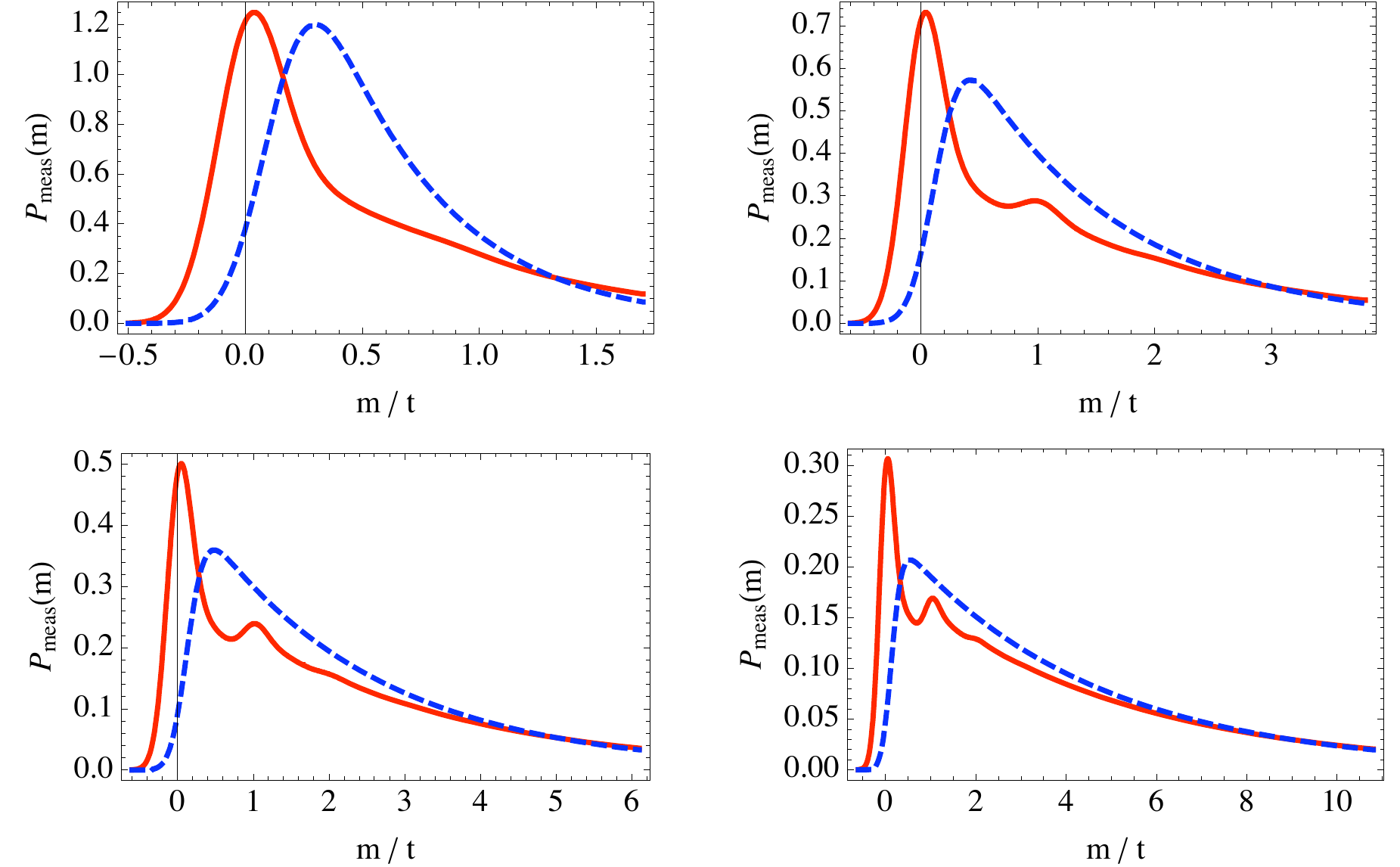}
\caption{
Distributions of the integrated output of the experiment, 
for both the cases of a classical oscillator 
($P_{\rm meas,cl}(m)$, dashed blue) and a quantum oscillator 
($P_{\rm meas,q}(m)$, solid red).  
Following Ref.~\cite{Harris08}, we have taken $\omega_M / 2 \pi = 10^5 {\rm Hz}$, $\gamma/ \omega_M = 1.2 \times 10^{-7}$ and $\Tbath = 300 {\rm mK}$.
In each case, we have assumed a noise level corresponding to $R=11$,
used an optimal averaging time, and have removed the zero-point shift in the average of the quantum distributions.  The panels correspond to initial oscillator temperatures of $\ninit=0$ (top left), $\ninit=1$
(top right), $\ninit=2$ (bottom left) and 
$\ninit=4$ (bottom right).   In each plot, the range of $m$ has been chosen to display $90\%$ of the 
area of the quantum curve.  At higher temperatures, multiple peaks are visible, however
the overall distinguishability from the classical distribution is reduced.}
\label{fig:ThermalR11DistPlots}
\end{center}
\end{figure}

In the left panel of figure \ref{fig:ThermalSNR11IIPlot}, we consider a situation where the added noise corresponds to $R = 11$, and plot the distinguishability between $\Pmeasq$ and $\Pmeascl$ (measured via $\II$) as a function of the initial oscillator temperature $n_{\rm init}$.  For each point, we have used an optimal averaging time; the dependence of this optimal time on $n_{\rm init}$ is shown in the right panel.  As in previous plots, we have also shifted the quantum distributions so that they have the same means as the corresponding classical distributions.  As could be expected, as the initial temperature increases, the distinguishability between classical and quantum distributions does indeed decrease.  However, this decrease is slow enough that one could obtain useful information even if the oscillator starts at low but non-zero temperature.  It is also interesting to note that while the overall distinguishability between the classical and quantum distributions decreases with initial temperature, the quantum distribution will develop multiple peaks;  this is shown in figure \ref{fig:ThermalR11DistPlots}.

\clearpage

%REFERENCES_______________________________________________________________________

\clearpage

\section*{References}
\bibliographystyle{unsrt.bst}
\bibliography{NJParXiv}

\end{document}